\documentstyle[amsfonts,epsf,,aps]{revtex}
\newcommand{\ee}{\end{equation}}
\newcommand{\be}{\begin{equation}}
\begin{document}

\title{Chromatin: a tunable spring 
at work inside chromosomes}

\author{Eli Ben-Ha\"{\i}m, Annick Lesne and Jean-Marc Victor\footnote{
Email adresses for correspondence: lesne@lptl.jussieu.fr, victor@lptl.jussieu.fr}} 
  
\address{
Laboratoire de Physique Th\'eorique
des Liquides, \\
Universit\'e Pierre et Marie Curie,  \\
Case courrier 121, 4 Place Jussieu, 75252
 Paris Cedex 05,
France\\
}

\date{\today}
\maketitle

\begin{abstract}
{\small
This paper focuses on mechanical aspects of chromatin biological functioning.
Within a basic geometric modeling of the chromatin assembly,
 we give  for the  first time the complete set of elastic constants
(twist and bend persistence lengths, stretch modulus and twist-stretch coupling
constant)  of the so-called 30-nm chromatin fiber, in terms of DNA elastic
 properties and geometric properties of the fiber assembly.
The computation naturally embeds the fiber within a current analytical model
known as the ``extensible worm-like rope'', allowing a straightforward 
prediction of the force-extension curves.
We show that these elastic  constants are strongly  sensitive to the linker
 length, up to 1 bp, or equivalently to its twist, and might locally reach 
very low values, yielding a highly flexible and extensible domain in the fiber.
In particular, the twist-stretch coupling constant, reflecting the chirality 
of the chromatin fiber, exhibits steep variations and sign changes when the 
linker length is varied.
 We argue that this tunable elasticity might be a key feature for chromatin
 function, for instance in the initiation and regulation of transcription.
}
\end{abstract}

\vskip 10mm\noindent
{\it PACS numbers: 82.37.Rs, 46.25.-y, 87.19.Rr, 83.80.Lz}
\twocolumn

\section{Introduction}

Chromatin is an ubiquitous protein-DNA complex observed in chromosomes of all 
eukaryotic organisms 
and strikingly conserved during evolution\cite{holde} \cite{Wolffe}\cite{Saen}.
It ensures DNA compaction
during the mitosis and plays a key but still unclear  role
in all the
biological functions involving genomic DNA: replication, transcription and repair.
For instance, the chromatin degree of compaction is acknowledged to regulate, somehow,
transcriptionally active regions\cite{widom2}.

The general issue  taken up in  this paper  is to understand 
the interplay between the  mechanical properties of the fiber and its biological
functions. We aim at understanding quantitatively the 
grounds for existence of the so universal and
so conserved features of the assembly of linkers and nucleosomes forming the
chromatin fiber.
We claim that part of the answer lays in the peculiar mechanical properties of this
assembly.

\vskip 3mm

The typical length scale of chromatin fiber (diameter about 30 nm)
is large enough to allow a mechanistic viewpoint: describing the elastic properties of
the chromatin fiber is nothing but a problem of spring mechanics \cite{love}.
Nevertheless, the architecture of this ``spring'' is much more complex than a simple
helical coiling and we expect that the detailed structural features of the chromatin 
assembly still strongly influence the  behavior at the fiber scale.
 We  thus investigate the specific 
elastic behavior associated with the
chromatin structure and its sensitivity to the structural parameters of the fiber,
singled out within a basic geometric modeling of its assembly.

\vskip 3mm
We underline that fiber elasticity is involved in two, quite different, issues.
A first one is to describe the linear elastic response of the fiber to global
stresses, i.e. a force and a torque applied at its ends.
This issue refers to micromanipulations in which a single chromatin fiber
is pulled (and possibly will be twisted) \cite{cui-bust}.
Our theoretical study provides a framework to interpret 
the experimental results, in particular 
 to predict force-extension curves in terms of geometric and
mechanical parameters of the underlying model of fiber.
Our approach largely extends preliminary results presented quite recently by
Schiessel et al.\cite{schiessel}, since it gives the exact analytical value
of the complete set of elastic constants of the fiber.
Comparison with the observed curves allows to extract small scale information, either
about the microscopic structure, either about the interactions involved, either about
possible conformational or structural   changes.

However, except during anaphase, when sister chromatids are
separated by the mitotic spindle \cite{duplantier}, 
such ``macroscopic'' stresses are not encountered {\it in vivo} at the chromatin fiber
level;
micromanipulations of the fiber and associated force measurements 
may nevertheless  unravel physical parameters involved
in  biological mechanisms, such as the elastic constants.
An issue directly relevant to the {\it in vivo}
functioning of chromatin is  to describe the response 
of the fiber to local, internal stresses as
those created by intercalators, groove-binding proteins, or any induced change in the
fiber assembly or small-scale structure.
Our theoretical approach also gives a framework to such studies.
Indeed, to be solved, both issues require to relate the structure and mechanics
at the DNA scale and those at  fiber scale, which is the scope of the present work.

\vskip 3mm
The chromatin scale is precisely  the scale of nanomechanics:
at this scale, we expect a strong and direct interplay between the biological
functioning, monitored  by various enzymes, and the mechanical
properties of the substrate, here the chromatin fiber.
We thus believe that a mechanistic approach is  well-suited 
 to evidence possible mechanisms 
for the fiber  decondensation prior to transcription, 
for the connection between enhancer and promoter DNA regions
during transcription, 
for the nucleosomal DNA site exposure allowing protein binding
at specific sites or for the ejection of 
nucleosomes presumably required during replication\cite{revuewidom}.

\vskip 3mm
The core of the study
 is to describe how the linear elastic behavior of chromatin fiber
 originates from the elastic properties of linker DNA
(i.e. naked DNA connecting the nucleosomes).
In Section II, we introduce a mechanical model of chromatin
fiber, quite similar to the so-called two-angle model\cite{struct-4}; it
incorporates
microscopic data such as  DNA structure and elastic
properties \cite{croquette}\cite{Bouchiat}  as well as  nucleosome cristallographic
 data \cite{Luger} into an analytically tractable assembly.
Its implementation  gives the geometric properties of the chromatin fiber, 
presented in Section III with a special mention to symmetry properties and to
 the quantities
relevant to the mechanics of the fiber.
Section IV is devoted to the elastic properties of the fiber. 
The first step of their study is to  
  relate  the distribution of stresses  along linker DNA to
the ``macroscopic'' stresses (force and torque) applied to the fiber ends.
This allows to compute analytically
 the elastic constants describing  the
linear response of the fiber from the knowledge of naked DNA elastic constants, and to
investigate quantitatively how they vary with the relaxed fiber structure, itself
controlled by the ``microscopic'' structural parameters,
as linker length $l$,
or equivalently its relaxed twist $\tau^0$,
 and entry/exit angle $\Phi$ of linker DNA on the nucleosome \cite{struct-4}.
The results  substantiate an effective continuous description of the 30-nm
fiber as an extensible worm-like rope (EWLR) extending the 
classical worm-like chain model\cite{WLC} to an extensible and twistable chain
\cite{marko-stretch-1}\cite{marko-stretch-2}\cite{kamien}. 
In Section V, numerical results are presented  and their physical meaning
and implications are  discussed.
We then enlight our analytical results
by comparing them  with experimental results
obtained by Cui and Bustamante\cite{cui-bust} by pulling a single chromatin fiber.
Section VI presents  the biological relevance of our study. 
In particular, we
 exploit our complete and quantitative analysis
to discuss how the elastic properties of the chromatin fiber
might at the same time
 favour DNA compaction into the
chromosomes and 
 allow 
local decondensation of chromatin involved in gene expression.

\section{Assembly of a model chromatin fiber}

Chromatin is composed of a double-stranded DNA molecule wrapped from place to
place around histone cores
\cite{Saen}.
We here focus on the fixed-nucleosome case, corresponding to
deacetylated histone tails\cite{revuewidom}.
This allows to  
 model separately the nucleosomes and the linkers, i.e.
the naked-DNA segments connecting the successive nucleosomes.
Our model amounts to build step-by-step, i.e. nucleosome after nucleosome, a chromatin
fiber. It  
incorporates aknowledged data on  DNA structure, its bend 
and twist persistence lengths, and on the nucleosome structure.
Each step of the assembly is described analytically, but  
we implemented this model within a Maple program in order to handle an arbitrary number
of nucleosomes and to perform a quantitative analysis of the SH geometric properties.

\subsection{Linker modeling}

Linker DNA is in the usual B-DNA form: it is a right-handed double-helix, of radius 
$a\approx 1$ nm and pitch equal to $l^0=$ 3.4 nm , which corresponds to $n_{bp}^0=10.6$
base pairs  (bp) per turn.
We shall suppose in the following  that all the linkers have the same
number  $n_{bp}$  of base pairs,
i.e. the same length $l=n_{bp}l^0/n_{bp}^0$.
It corresponds {\it in vivo} to phased nucleosomes 
observed  for instance in satellite chromatine\cite{Widomlowary}.
It corresponds {\it in vitro} to reconstituted fibers, involving repeated sequences 
each having a strong positioning effect on
nucleosomes\cite{bednar}\cite{definedarrays}.
In fact, we shall need this hypothesis to be satisfied only  locally, over a
few linkers, so that our model also accounts for  irregular
native chromatin.

Since we are looking for generic properties of the chromatin fiber, originating from
its assembly, sequence effects are ignored
(besides,  they
might be treated  in a second stage, within the same modeling, as
 local strains superimposed to the relaxed
homogeneous structure that we here consider).
Without sequence effects, the linker is straight in absence of applied
constraints;
experimental evidence for straight linkers in relaxed fibers supports
our generic  modeling \cite{struct-holde}.

Linker DNA will be considered as an homogeneous cylindrical rope of radius $a$, that
may twist
and bend, but not  stretch: 
due to the large value of the stretch modulus $\gamma_{DNA}\approx 1200$ pN
\cite{marko-stretch-1}\cite{coeff-g},
stretch energy and strain can be ignored in linker DNA, at least in the 
low-stress situations that will be considered here
(forces below 20 pN).
Linker DNA is thus 
seen  as a non extensible semi-flexible polymer
and described within the continuous worm-like rope model
(worm-like chain model supplemented with twist energy \cite{Bouchiat}); 
the elastic energy densities (energies per unit length) thus  write:
\be
\begin{array}{rll}
\epsilon_{twist}&={k_BTC\over 2}(\omega-\omega^0)^2
&\hskip 10mm
C= 75 \;{\rm nm}
\hskip 3mm({\rm twist})
\\ && \\
\epsilon_{bend}&={k_BTA\over 2}\;\rho_{DNA}^2&\hskip 10mm
A= 53 \;{\rm nm}\hskip 3mm({\rm bend})
\end{array}
\ee
where $T$ is the temperature, $k_B$ the Boltzmann constant,
$\rho_{DNA}$ the local curvature  of the constrained linker and $\omega$ the twist
rate. The relaxed twist rate $\omega^0=2\pi/l^0$
 is
 supposed to be homogeneous, which again amounts to ignore sequence effects.
The constants $A$ and $C$ are respectively the bend and twist persistence lengths
of DNA, here given for 10mM NaCl (or any other monovalent
salt)\cite{kam}\cite{croquette}\cite{Bouchiat}.
We expect $C$ to be almost independent of the ionic strength, since twist constraints
are not strongly affected by electrostatics. By contrast, the electrostatic
contribution to $A$ is important\cite{odijk}\cite{netzorland}.
 The non-electrostatic contribution 
 provides a lower  bound  $A\geq 23$ nm, which is in fact an overestimate;
the experimental lower bound is rather $A\geq 40$ nm \cite{salt}.
\vskip 3mm
We choose  a direction along the DNA path corresponding 
to increasing  indices $j$.
We denote  $s\in[0,l]$  the arclength along the dsDNA axis.
Geometrically, the linker $j$ is described 
 as represented on Figure 1 by the local axis $\vec{u}_j(s)$
of the double helix 
and a vector $\vec{t}_j(s)$, locally tangent to the minor groove
and rotating around $\vec{u}_j(s)$ with an angular rate $\omega$.
 $\vec{t}_j(s)$
accounts for the  torsional state of the linker:
the  angle of the rotation transforming $\vec{t}(0)$ into $\vec{t}(l)$
 is precisely the twist
$\tau=l\omega$ of the linker. The vectors
$\vec{u}_j(s)$ and $\vec{t}_j(s)$ are normalized to 1 and make a constant angle:
$\vec{u}_j(s).\vec{t}_j(s)=\sin(\alpha_{DNA})$ independent of $j$ and $s$.
The slope of the strands in the double helix writes
$\tan(\alpha_{DNA})=l^0/2\pi a$, hence $\alpha_{DNA}\approx 28^o
$. The vector 
$\vec{u}_j(s)$ is independent of $s$ in the relaxed fiber
(a variation with $s$ would describe the linker bending in  constrained state),
whereas $\vec{t}_j(s)$ rotates around $\vec{u}_j$ with the relaxed twist rate
$\omega^0$.

\subsection{Nucleosome modeling and  assembly of the chromatin fiber}
Nucleosome
structure is now well-known, thanks to high resolution
cristallography data\cite{Luger}.
Histone tails lock DNA on the histone core, so that
nucleosomes are  fixed on the DNA and 
 the net
effect of the nucleosome on the DNA trail can be described as a rigid 
 kink connecting the linkers $j$ and $j+1$. In consequence, twisting the linker $j$ of a given  angle
reflects in a rotation of the same angle  of 
the linker $j+1$ around
the linker $j$.
The rotational positioning of nucleosome $j+1$ with respect  to
nucleosome $j$ is thus entirely prescribed by the twist angle $\tau=l\omega$.
In the relaxed state, it is equivalently prescribed  by
the linker length $l$  ($\omega=\omega^0$,
$\tau^0=l\omega^0$).

\vskip 3mm
 Technicalities about the assembly and
its geometric description  are presented on Figure 2.
The linker $j$ enters the nucleosome $j$ at an 
entry  point $E_j$ well-localized on the histone core. 
The linker DNA is grafted to the nucleosome by its 
minor groove\cite{Luger}; this  implies that the
 plane spanned by $\vec{u}_j$ and
$\vec{t}_j(l)$ prescribes the position and the orientation of the nucleosome, 
given its radius 
$r_{nucl}=5.5$ nm, its height $H=5.7$ nm
and the angle $\alpha=4.47^o$ between the 
linker and
 the nucleosome axis $\vec{N}_j$ ($\vec{u}_j.\vec{N}=\sin\alpha$).
As shown in Figure 2 (right), $\vec{u}_j$, $\vec{t}_j(l)$ and $\vec{N}_j$
belong to the same plane, which is nothing but 
  the tangent
plane to the nucleosome at $E_j$.
Similarly, the plane tangent to the nucleosome $j$ at the exit 
point $S_j$ where linker $j+1$ leaves the nucleosome is spanned
by $\vec{u}_{j+1}$ and
$\vec{t}_{j+1}(0)$, which thoroughly determines
this outgoing linker.
The relative positioning of the linkers is  conveniently described
by the angle $\Phi$ ($\Phi>0$ by convention)
between their projections
on a plane
perpendicular to the nucleosome axis  $\vec{N}_j$.
The structural effect of nucleosome
$j$ amounts to a  translation in space  from  the entry point
$E_j$  to the exit
point $S_j$ and a rotation of an angle $4\pi-\Phi$ around 
the nucleosome axis $\vec{N}_j$, transforming $(\vec{u}_{j}, \vec{t}_{j}(l))$ into 
 $(\vec{u}_{j+1}, \vec{t}_{j+1}(0))$.
As long as we  consider 
an homogeneous fiber, $\Phi$ is independent of $j$.
The line passing through the nucleosome center $G_j$ and directed along the bissector
$\vec{D}_j$ of
this angle $\Phi$ is the so-called dyad axis of the nucleosome $j$.
Experimental evidence that the entry-exit angle $\Phi$ varies with salt concentration
(it contracts when salt concentration increases)
\cite{bednar}\cite{struct-holde}\cite{leuba}\cite{leuba-H1}
 motivates us to study various values of $\Phi$.
We here ignore interactions between nucleosomes. This is valid in low-salt
conditions, in which the chromatin is expected to be extended, with internucleosomal
distances larger than the interaction range.
In high-salt conditions, our model aims at describing the ``geometric'' contribution
to chromatin elastic behavior,  i.e. the contribution originating from the DNA elastic
properties and relayed by the chromatin fiber architecture.
Comparison with observed elasticity would give access  to the neglected interactions.

\subsection{Linker histones and chromatosome modeling}
{\it In vivo}  and {\it in vitro} experiments show that the condensed
conformations of the chromatin fiber are obtained for a more complex assembly of
DNA and histones: an additional histone H1 (or a close chemical variant H5, for
instance
in chicken erythrocytes)
is bound to linker DNA near its entry/exit site on the
nucleosome\cite{struct-holde}\cite{H1-thomas}.
Presumably, the role of this ``linker histone'' is both to secure the wrapping of DNA
around the nucleosome and to stabilize the DNA helical coiling forming the 30-nm fiber
\cite{bednar-H1}\cite{H1-holde}.
Digestion experiments evidenced that  the  ``core particle''
 now involves 166-168 bp of DNA (among which the 146 bp wrapped around the histone
octamer); this new, larger,  entity is called a chromatosome.
The position of linker histone with respect to linker DNA and nucleosome,
as well as the wrapping of DNA around it, are still debated
\cite{H1-thomas}\cite{H1-holde}.
It is only recognized that 166-168 bp are protected and that the angle
(still denoted $\Phi$) 
between ingoing and
outgoing linker DNA is lowered,
varying with
ionic strength.
Our model easily accomodates the presence of linker histones along the fiber since
only the resulting kink of  DNA path is to be described.
 We may roughly account for the presence of linker histone by modifying
$\Phi=130^o$ (without linker histone) into $\Phi=90^o$
at low salt down to $\Phi=45^o$ in high salt\cite{leuba-H1}\cite{H1-holde}; 
the nucleosome is then replaced, within the same two-angle modeling,  by an
effective cylindrical chromatosome, with possibly different values for $r_{nucl}$,
$H$ and $\alpha$. 
For instance $r_{nucl}=0$ and a lower value of $H$ would  reproduce the crossing of
 ingoing and
outgoing linkers in the neighborhood of the linker histone.
Also, the effective length of the linkers is to be reduced by 20-22 bp.

We might thus  study within the same geometrical modeling  all the different
instances encountered for the nucleosome: basic nucleosome or nucleosome dressed with
H1, with or without chemical modification of H1 tails.
The specific biological details (H1 positioning, possible acetylation of histone
tails \cite{leuba-acetyl}) would be taken into account first in $\Phi$, whose value is experimentally
accessible, then in a second step by a precise fit of 
 the parameters $r_{nucl}$,
$H$ and $\alpha$ of the effective cylindrical ``core particle'', so as to
reproduce accurately  the kink induced in the linker DNA broken line.
Nevertheless, a systematic study led us to think that $\Phi$ is a far more important control
parameter than  $r_{nucl}$,
$H$ and $\alpha$, which will be  henceforth considered as fixed characteristics.


\vskip 5mm
To summarize, the parameters of the model, to be varied in our study, 
 are the two angles $\tau^0$  (or equivalently $l=\tau^0/\omega^0$ or $n_{bp}$) 
and $\Phi$.
 The angle $\tau^0$ determines the relative positioning of two successive
nucleosomes, whereas the angle $\Phi$ determines the relative positioning of two
successive linkers.

\section{The chromatin superhelix (SH)}

The regularity of the microscopic assembly
 enforces an helical organization:
we shall speak
of the chromatin superhelix (SH).
We now describe the main geometric properties of this 
helical coiling resulting from the  regular assembly of linkers and 
nucleosomes.
As shown in Figure 3, our qualitative structural predictions recover standard structures proposed yet long
ago\cite{struct-1}\cite{struct-2}.
In  particular, we note that the various structures obtained when varying $\tau^0$ 
(or equivalently $n_{bp}$) and
$\Phi$ cover both ribbon structures (less than 3
nucleosomes per turn) and cross-linked structures (more than 3 nucleosomes per turn).
We avoid to speak of ``solenoid'' structure as this term refers to a model with bent
linkers\cite{solenoide}, whereas our modeling considers that linkers are straight in the relaxed fiber.

\subsection{Symmetry properties of the SH}

We first underline the symmetry properties of the SH, of much importance since they
will reflect in the elastic properties of the fiber in the linear regime.
The SH exhibits a discrete helical invariance:
all the linkers (respectively all the nucleosomes) are equivalent,
up to a rotation of angle $\theta$ around the axis $\vec{A}$ of the SH and a translation along this axis, 
bringing the linker $j$ onto the linker $j+1$ (respectively the 
nucleosome $j$ onto the nucleosome $j+1$).
The nucleosome centers all lay on a cylinder of axis $\vec{A}$
and radius $R$, whereas the entry and exit points all lay on a cylinder of radius $r$
(see Figure 4).

As the two strands of dsDNA are oriented in opposite directions
(see Figure 1), 
there is no privileged orientation along linker DNA and 
only the nucleosome
geometry can break the symmetry upon reversal of the fiber.
Hence the  fiber is not oriented provided the entry and
exit points in a nucleosome have the same characteristics: 
it is the case in absence of linker histone, or when linker histone is positioned
symetrically with respect to $E_j$ and $S_j$.
The fiber properties are then unchanged under the transformation:
\be
\begin{array}{rclrcl}
j&\longleftrightarrow&N+1-j&\vec{N}_j&\longleftrightarrow&-\vec{N}_{N+1-j} \\
S_j&\longleftrightarrow&E_{N+1-j}&\vec{A}&\longleftrightarrow&-\vec{A} \\
\vec{u}_j&\longleftrightarrow&-\vec{u}_{N+1-j}\hspace{15mm}&\hspace{15mm}P_j(s)&\longleftrightarrow& P_{N+1-j}(l-s)
\end{array}
\ee
Here $j=0\ldots N+1$ where $N$ is the number of nucleosomes in the  fiber
beginning in $S_0$ with  linker 1 and ending with  linker $N+1$, at $E_{N+1}$
It follows  that the nucleosome  axes
are all tangent to the cylinder containing the nucleosome centers, 
and  that the line $\vec{D}_j$ relating the nucleosome center $G_j$ to its 
orthogonal projection onto
the SH axis is nothing but the dyad axis of the nucleosome $j$
(see Figure 5, left).
This dyad axis is invariant under the above reversal transformation.
 When $\tau^0$ is varied, the linker direction $\vec{u}_j$ sweeps a cone of axis
$\vec{D}_{j}$ (or $\vec{D}_{j-1}$, due to the symmetry upon
reversal of the SH) and angle $\xi$, depending on $\Phi$ (see Figure 5, right). 
This property simply reflects the fact that the geometry of the fiber assembly is
thoroughly determined once the nucleosome is positioned with respect to the SH axis,
due to the symmetry properties and to the fact that the junction between the nucleosome
and the ingoing and outgoing linkers is a rigid kink.
Indeed, by symmetry, the nucleosome dyad axis $\vec{D}_j$ is orthogonal to the SH axis
$\vec{A}$ and points towards it; the remaining ``degree of freedom'' is the orientation
of the nucleosome axis $\vec{N}_j$ in the plane orthogonal to $\vec{D}_j$;
when $l$ varies, $\vec{N}_j$ rotates together with $\vec{u}_{j}$ and $\vec{u}_{j}+1$
around $\vec{D}_j$.
This orientation is  thoroughly described by the angle $\beta\in[0,2\pi[$ 
between its axis $\vec{N}_j$ and the SH axis $\vec{A}$,
in the frame $(\vec{A},\vec{D}, \vec{A}\wedge \vec{D})$. 
This angle satisfies $\cos\beta=\vec{N}_j.\vec{A}$.

\subsection{SH geometric characteristics}
 The relevant quantities to be computed (lengths $D$ and $r$,
angles $\theta$, $\beta$,  $z$, $\eta$, and $\xi$)  are
shown on Figures 4, 5 and 6.
All these quantities are independent of $j$, due to the (discrete) rotational invariance of the
SH.
When linker lengths are all increased by the DNA pitch $l^0$, the twist angle of each
linker increases by $2\pi$, hence the relative orientations of 
successive linkers, or successive  nucleosomes, 
are unchanged.
 It follows that the SH shape is preserved: 
the angles (as $z$, $\eta$ or
$\theta$) are unchanged whereas the length $r$
is scaled  by a factor
$(1+l^0/l)$  (when
comparing $l$ and $l+l^0$);  the lengths $R_{SH}$ and  $D$  are also scaled, but the 
involved scaling factor is smaller, since the
contribution of the nucleosomes to the actual size of the SH is the same
for $l$ and $l+l^0$; the precise value of this  factor moreover depends
on the structure of the SH (nucleosome orientation, for instance). 

 We here underline two points.
First, the orientation of the SH axis is chosen so that 
the 
distance $D$ between two successive  nucleosomes along the SH axis $\vec{A}$
is positive:  $D>0$ 
(i.e. $D=\vec{A}.[\vec{G_jG_{j+1}}]>0$ by definition of $\vec{A}$).
Also,  since the structure is discrete, the angle
$\theta$
of the rotation ${\cal R}_{\theta}$
around the SH axis $\vec{A}$, transforming the projection of a nucleosome
onto the plane orthogonal to $\vec{A}$ into the following one
 is not univoquely defined:
 $\theta_0>0$ and $\theta_0
-2\pi<0$ are both possible (think to the array of grains in a corn ear or to the cells
on a pineapple peel). 
By convention, we choose the
value of smallest modulus, hence  let
$\theta$ vary in $]-\pi, \pi]$; it corresponds to the (projected) angle swept by the
geodesic (the array of grains of largest slope)
 relating $G_j$ to $G_{j+1}$ onto the cylinder of axis $\vec{A}$ and radius
$R$.

$\Delta$= dist$(G_1,G_2)$ and the angle $\Psi=\widehat{G_1G_2G_3}$  defined by three
 The pitch $P$ of the SH is given by $P=2\pi D/|\theta|$.
Note that the choice of the value of smallest modulus for $\theta$ is essential to get
the actual value of the pitch through this formula. 
The degree of compaction can be measured as 
the number of nucleosomes per 10 nm of fiber, equal to $10/D$
when $D$ is given in nm. 

\subsection{Excluded-volume effects}
Excluded volume (of nucleosomes but also linkers) has to be taken into account
to discard unrealistic structures.
In consequence,
the above results should be supplemented by a geometric condition assessing whether
they actually correspond to a  possible structure, 
i.e. compatible with excluded-volume
constraints.
We here rather state an upper bound, involving only the geometric parameters of the
fiber.
Recall that $\beta$ is the angle between the nucleosome axes 
$\vec{N}_j$ and the SH axis (independent
of $j$ by symmetry).
A sufficient condition ensuring that no steric hindrance
is encountered in the SH structure is that $P>P_c$ where: 
\be\label{pitchbound}
P_c= H\cos\beta +2R|\sin\beta|\ee
where $R$ is the distance of the nucleosome centers to the SH axis (computed in the
Maple implementation) and $H$ the height of a nucleosome. For $\beta=0$, 
this upper bound gives the exact threshold
$P_c=H$.
All linker lengths such that 
$P(l,\Phi)\leq P_c$ are forbidden. Note that a local 
breaking of this criterion is allowed:  it is still  possible
 that a few linkers lengths (less than 
$2\pi/\theta$) take  this value.
A more accurate check can be obtained by drawing
for each configuration the developped representation of the
SH, i.e. the mapping on a plane of the fiber track on the cylinder of radius $R$,
as shown on Figure 7.
Its construction is straightforward, knowing  the 
geometric characteristics $D$, $\theta$ and $\beta$. Excluded-volume constraints
are satisfied if the tracks of any two nucleosomes do not intersect.   
This method requires to check each configuration, hence comes after the criterion on
$P-P_c$, when it fails.

\subsection{Numerical results}

The major interest  of our numerical implementation
is to go further than a qualitative description and to 
compute explicitly any  geometric  characteristic of the relaxed fiber,
in order  to study quantitatively its variation with $l$ (or $\tau^0$)
and $\Phi$.
The main results are presented in Figures  8, 9 and 10 for a linker length between 30
and 50 bp (usual case).
The quantitative geometric characterization of the SH is essential to compare
the different structures obtained for different values of $(l, \Phi)$.
It is also interesting from an experimental viewpoint, since
it would allow to extract informations on the chromatin assembly from the
experimentally accessible values of $D$ and $\theta$.

We  first evidence that a regular helical packing of nucleosomes
is compatible with straight linkers and 
does not require nucleosome-nucleosome interactions (but the
special configurations exhibiting columnar arrays of nucleosomes
are likely to  be stabilized by such internucleosomal
interactions). Mainly, our systematic analysis shows that it always  
lead to a 30-nm fiber.
 Although the detailed structure of the fiber  is strongly sensitive to 
linker length $l$
 and entry/exit angle $\Phi$,
the
 value of about 30 nm for the fiber
diameter $2R_{SH}$ is a robust feature, hence of low structural significance;
 for instance, it
does not discriminate close cross-linked structures and more extended ribbons;

We enlight some remarkable geometries (see also Section V.E):

\noindent $\bullet$ \
 a ribbon -- or zigzag -- structure is obtained  for $\theta=\pi$,
when  the number of nucleosomes
per SH turn reaches the minimal value 2; then all the nucleosome dyad axes are aligned,
$\beta$ is close to $\pi/2$ but slightly different: $\beta=\pi/2-\alpha$
 (see Figure 14).

\noindent $\bullet$ \ a particular packing is obtained when $n_{bp}/n_{bp}^0$
 equals an integer;
 the nucleosome axes are then
all aligned, hence all parallel to the SH axis: $\beta=0$.
A fine tuning of $n_{bp}$ or $\Phi$ brings then into columns as seen on Figure 3.

\noindent $\bullet$ \ the vanishing of the pitch  $P$ is accompanied by a reversal of the SH axis,
from which follows that  $\cos z$ exhibits a jump from a value $c$ to the value
$-c$ and $\theta$ jumps to the value $2\pi -\theta$.
We shall see that the vanishing of the pitch $P$ is an important (although virtual)
 event, having 
 striking implications on the elastic properties of the fiber.

\noindent $\bullet$ \ a change of chirality is observed for $\eta=\pi$: the SH is right-handed for
$\eta\in[0,\pi]$ and left-handed for $\eta\in[\pi,2\pi]$. For $\eta=\pi$, linkers
cross the SH axis.


One has to carefully distinguish between:

\noindent
-- the handedness of the track passing through the nucleosome centers.
As we have  seen above in the definition of $\theta$, the handedness of this discrete
structure is ill-defined. A same architecture can be seen as right-handed (direct rotation
${\cal R}_{\theta}$) or left-handed
(indirect rotation
${\cal R}_{\theta-2\pi}$).
The choice of the value of smallest modulus, adopted here, corresponds to 
the shortest path. This ``pseudo-chirality'' is then given
by the sign of $\theta$  and changes  
when $\theta$ crosses the value $\pm \pi$ or 
when $D$ (or equivalently the pitch $P$) vanishes.
We underline that this structural feature does not define a relevant chirality, due
to the arbitrariness of its definition, and should not be confused with the chirality
of the SH, defined as follows;

\noindent
-- the chirality of the  linker DNA trail (a broken line);
it is determined by the position of $\eta$ with
respect to $\pi$, i.e. by the sign of $\cos(\eta/2)$.
As fiber elasticity originates from linker elasticity, this chirality is the
only  relevant
one for our mechanical study and will be henceforth adopted
as the definition of the fiber chirality.
We expect this chirality to determine the coupling between the twist and stretch
elastic degrees of freedom i.e. the sign of $g$ to coincide with the chirality. 

In the same spirit, we warn about the difference between the fiber 
and an helical DNA coiling passing through the nucleosome centers, in particular 
in what concerns their mechanical properties.

\vskip 3mm


One of the conclusions of this thorough structural study is the fact 
that the connection  between the microscopic parameters and the SH geometric
characteristics  is too complex and multivariate to get 
small-scale informations from structural
observations of the fiber.
We underline that the study of chromatin structure is not
in itself  sufficient
to unravel its biological functioning, all the more as  structural observation
of the fiber is difficult and some results  questionable\cite{revue}.
Moreover, structural changes might not be the only way to pass local DNA modifications
to higher scales; the functioning might rather be controlled by elastic properties.

For these two reasons, we turn to the analysis
of the mechanical properties of the fiber, which are now at hand since
we know explicitly
 all the geometric characteristics of the fiber assembly.


\section{Analytic calculation  of the elastic constants of the chromatin fiber}

\subsection{Modeling the chromatin fiber as an extensible worm-like rope (EWLR)}

Our aim is to study the elastic response  of the fiber to external stresses
at scales larger than its pitch.
We thus consider that we apply to the fiber 
a force $\vec{F}$,  along the fiber axis $\vec{A}$; if $\vec{F}$ were not directed along the SH axis, it would induce a transitory motion,
compelling the axis to align itself with the force direction,
at least far from the ends. We here only suppose
that this constrained equilibrium is yet reached.
We may also apply a torsional torque $M_t$
(directed along $\vec{A}$) and a flexural torque $\vec{M}_b$,
i.e. a torque component orthogonal
to $\vec{A}$.
Computation of  elastic coefficients describing the linear response of the fiber
to the applied force and torque will be performed analytically within a continuous
description of the fiber, i.e. an effective large scale description in which the
discrete nature of the assembly is smoothed out.
We denote
$S$  the arclength along the axis of the relaxed SH,
$u(S)$ its local relative extension, 
$\Omega(S)$ its local twist rate and  $\varrho(S)$ its local curvature,
with respect
to the straight and untwisted  relaxed state of the fiber
for which $u$, $\Omega$ and $\rho$ thus identically vanish.
Due to the axial symmetry of the relaxed SH
at scales larger than its pitch, the bending energy density 
does  not depend on the direction of the bending so that it involves only 
the total curvature $\rho$ and the modulus $M_b$ of the flexural torque
(the special instance of ribbon-like configurations that break this axial symmetry
is discussed below in Section V-E).
The fiber has thus only  three 
degrees of freedom $u$, $\Omega$ and $\rho$, local along the fiber.
These  strains $u$, $\Omega$ and $\rho$ are the canonical variables of  the density
of elastic free energy, 
which reflects in the following differential form:
\be\label{diffepsilon}
d\epsilon_{SH}=F\;du+M_t\;d\Omega+M_b\;d\rho\ee
We restrict to the linear response regime, which expresses in the following relation:
\be\label{linearansatz}
\left(\begin{array}{l}
F\\M_t\\M_b
\end{array}\right)=
\left(
\begin{array}{ccc}
\gamma&k_BTg&0\\
k_BTg&k_BT{\cal C}&0\\
0&0&k_BT{\cal A}
\end{array}
\right)
\left(
\begin{array}{l}
u\\ \Omega\\\rho
\end{array}\right)\equiv \Gamma\left(
\begin{array}{l}
u\\ \Omega\\\rho
\end{array}\right)
\ee
where it can be shown that the stress-strain tensor $\Gamma$
is necessarily symmetric (a special instance of Onsager relations).
This linear  response ansatz relates the  strains $u$, $\Omega$ and $\rho$ of the fiber
and the  stresses $F$, $M_t$ and $M_b$  experienced by the fiber.
${\cal A}$ is the bend persistence length of the fiber, ${\cal C}$
its twist persistence length, $\gamma$ its stretch modulus
(dimension of a force) and $g$ the twist-stretch
coupling constant (no dimension).

Plugging the linear response ansatz (\ref{linearansatz}) into 
(\ref{diffepsilon}) leads to the SH
density
 of elastic free energy:
\be\label{EWLR1}
\epsilon_{SH}(S)={k_BT{\cal A}\varrho^2(S)\over 2}+{k_BT{\cal C}\Omega^2\over 2}+
{\gamma\,u^2(S)\over 2}
 +k_BTg\,\Omega(S)u(S)
\ee 
Such a continuous description
of the fiber can be termed ``extensible worm-like rope''
  model  (EWLR) \cite{marko-stretch-2}. It extends the WLC model
introduced in 1949 by Kratky and Porod and currently used to describe
stiff polymers\cite{WLC}
by accounting for  twist (as  above for linker DNA\cite{Bouchiat}) 
 but also stretch  degrees of freedom.
The fact that chromatin is chiral  demands 
a linear coupling between twisting and stretching.
Due to the axial symmetry of the SH, $(\vec{F},\vec{M}_t,
\vec{M}_b)$ and $(\vec{F},\vec{M}_t,
-\vec{M}_b)$ (and also $(\vec{F},\vec{M}_t,
{\cal R}\vec{M}_b)$ where ${\cal R}$ is any rotation around the axis ${\vec A}$
leaving the SH unchanged)
should induce the same energy change in the SH. 
This symmetry argument shows that there is no other 
 coupling term at the linear order considered here\cite{marko-stretch-1}.
This model has been fully investigated by many groups in the context of DNA;
in particular, force-extension curves have been obtained \cite{marko-stretch-2};
a nonlinear term $V(u)$ might be added in (\ref{EWLR1}) to go beyond the linear response
regime.
These results can be straightforwardly  transposed to chromatin fiber, so that the
description of the harmonic elastic behavior of the chromatin fiber reduces to the computation of
the four elastic coefficients ${\cal A}$, ${\cal C}$, $\gamma$ and $g$
involved in the EWLR model.
Their determination from the computation of the elastic energy stored in the 
constrained linkers will prove that the fiber actually fits in an
 EWLR model when considered at large  enough  scale.

\subsection{A general analytic  method for computing elastic coefficients}

The first aim of our study is to express the
elastic coefficients
${\cal A}$, ${\cal C}$, $\gamma$
and $g$ as a function of the elastic coefficients of
linker DNA, given  the relaxed geometry of the fiber. A key point of our approach is
to relate the {\it stresses} exerted respectively on the
linker, considered as a WLR, and on the fiber, considered as an EWLR
instead of relating the {\it strains} $(\omega-\omega^0, \rho_{DNA})$ of the linkers and the
{\it strains} ($u, \Omega, \rho$)
arising at the fiber level (this approach appears to be technically cumbersome and
analytically intractable, see Section VII.B).
Although the stresses $F$, $M_t$ and $M_b$ are not the canonical variables
of the
free energy density $\epsilon_{SH}$, we shall express it as a function of these
variables $F$, $M_t$ and $M_b$. 
Plugging (\ref{linearansatz}) into (\ref{EWLR1}) leads to the following expression of
the free energy density:
\be\label{EWLR2}
\epsilon_{SH}=
{[k_BT{\cal C}F^2+\gamma
M_t^2-2k_BTgFM_t]\over 2(k_BT{\cal C}\gamma-k_B^2T^2g^2)}\;+\;
{M_b^2\over 2k_BT{\cal A}}
\ee
The principle of the computation is the following:
knowing the relaxed geometry of the fiber, i.e. the quantities
$r$, $\eta$, $z$ and $D$ computed in Section III,
 it is possible to
analytically 
determine the local stresses experienced by  linker DNA
when the fiber is constrained at its ends (and only at its ends)
and to deduce the elastic energy of a linker as a quadratic function 
of $(F, M_t, M_b)$.
The identity of 
the elastic energy of the fiber and the elastic energy stored in the linkers
will allow a termwise identification with (\ref{EWLR2}),
which leads the values of ${\cal A}$,
${\cal C}$, $\gamma$  and $g$
as
analytical  functions of the geometric parameters of the fiber,
and through our numerical implementation, as a function of $l$ (or $\tau^0$)
and $\Phi$.

A much important intermediate result is to establish 
the relation between the local stresses 
(experienced by the linkers)
and the global stresses (applied on the chromatin fiber).
The current point on linker $j$, with arclength $s$, is denoted  $P_j(s)$.
Recall that linker $j$ leaves the nucleosome $j-1$ at $S_{j-1}$ and enters the
nucleosome $j$ at $E_j$ (see Figure 6).
We denote $\vec{f}_j(s)$ the force
and  $\vec{m}_j(s)$ the torque  exerted at the point $P_j(s)$ of linker $j$
by the upstream part of the fiber.
The main arguments involved in  the derivation are the following:

-- when only pulling the fiber, the  rotational invariance of the relaxed fiber
should be preserved when   end effects are ignored.
We  may (and in fact should) 
 focus on the universal, rotationally symmetric behaviour of the fiber,
observed far from its ends.
Indeed, the quadratic, rotationally invariant EWLR energy is to be fitted
only to the quadratic, rotationally invariant energetic contribution coming from the
linkers, otherwise the identification would not make sense.

-- we restrict to the linear regime. We may consider separatedly the different stresses applied to
the fiber and simply sum up their effects to recover the general solution.
The contribution 
of the applied torque $\vec{M}$ to  the local torque $\vec{m}_j(s)$
is merely  $\vec{M}$.

The conclusion follows using standard
 equilibrium equations of spring mechanics.
When external 
forces and torques are applied at the
ends only,  linear response hypothesis implies  that
the solution finally writes:
\vskip 4mm
\be\mbox{
\begin{tabular}{|rcl|}
\hline
&&\\
\hspace{10mm}$\vec{f}_j(s)$&=&$\vec{F}$\\&&\\
 $\vec{m}_j(s)$&=&$\vec{M}-[\vec{O_j(s) P_j(s) }]\wedge\vec{F}$\hspace{10mm}
\\&&\\
\hline
\end{tabular}}\ee
\vskip 4mm
where $O_j(s)$ denotes  the orthogonal projection of $P_j(s)$ onto the SH axis 
$\vec{A}$.
It gives the relation between the global stresses $\vec{F}$ and $\vec{M}$
exerted at the fiber ends and the local stresses experienced at the linker level, at
each point of the DNA path.
The rotational symmetry of the fiber 
ensures that $\vec{F}$ is directed along the SH axis $\vec{A}$ at
equilibrium. 

The term $[\vec{O_j(s) P_j(s) }]\wedge\vec{F}$ reflects the involvement of the fiber
architecture in the expression of the local torque.
As we restrict to the description of harmonic elasticity, the coefficients of 
$\vec{F}$ and $\vec{M}$ in the elastic energies of the linker will be computed within the
relaxed SH.
{\it We underline that we do not need to compute
the constrained shape of the linkers to describe the linear response
of the fiber to applied force and torque.}
We carry on the computation by exploiting the fact that  the fiber is not oriented.
It follows that:
$||\vec{O(E_j)E_j}||=||\vec{O(S_j)S_j}||=r$
and:
\be ||[\vec{O_{j}(s)P_{j}(s)}]\wedge\vec{A}||^2=
||\vec{O_{j}(s)P_{j}(s)}||^2=
r^2[\cos^2(\eta/2)\cos^2z+\sin^2(\eta/2)(1-2s/l)^2]\ee
where  $\eta$, $r$ and $z$ are the values associated to the relaxed SH.

\subsection{Small-scale grounds for  the EWLR modeling of the chromatin fiber}
Linear response ansatz applies both at the linker level (WLR model, with parameters
$A$ and $C$)
and at the SH level (EWLR model with parameters ${\cal A}$, ${\cal C}$, $\gamma$ and
$g$).
Knowing the stresses thus gives the strains, respectively $(\rho_{DNA},
\omega-\omega^0)$ and $(u,\Omega,\rho)$, if required, and the elastic energies. The
computation of the SH elastic constants rests on the obvious but essential fact that
the elastic energy stored in the SH, expressed by (\ref{EWLR2}) at the SH level, 
is nothing but the sum of the elastic energies stored in its linkers, since no
interactions are involved in our modeling.
\vskip 3mm
A key point is the different decompositions of the torques
into torsional and flexural components  
 at the fiber level and at the DNA level:
\be
\vec{M}(S)=M_t\vec{A}+\vec{M}_b(S)\;\;\;\;({\rm SH})\hspace{25mm}
\vec{m}(s)=m_{t}\vec{u}+\vec{m}_{b}(s)\;\;\;\;({\rm  linker})
\ee
 The relation between the two decompositions involves the fiber geometry.
We underline once more that in the linear response regime, only 
the relaxed geometry is involved.
The first decomposition gives the components involved in the EWLR energy density
whereas the second is required to compute the elastic energy of a linker.
Indeed, the elastic energy densities of a linker can be expressed as a
function  of local stresses, according to:
\be\label{linearDNA}
\left.
\begin{array}{l}
m_t=k_BTC(\omega-\omega^0)\\
m_b=k_BTA\;\rho_{DNA}
 \end{array}\right\}
\Longrightarrow 
\epsilon_{twist}={m_t^2\over 2k_BTC}
\hskip 5mm{\rm and}\hskip 5mm
\epsilon_{bend}={m_b^2\over 2k_BTA}
\ee
According to a general result\cite{love}, besides easy to check directly here, $m_t$ is constant along the linker (i.e.
independent of $s$) whereas $\vec{m}_b$ may vary with $s$ both in direction and
modulus.

The EWLR model is a continuous model. It makes sense to describe the elastic behavior
of the discrete chromatin structure by means of an EWLR model only at large enough
scale,
so that the discrete effects are smoothed out.
At lower scale, the specific orientation of each linker influences its  energy,
more precisely the contribution coming from the global stress $\vec{M}_b$, which
breaks the rotational invariance;
the persistence
length of bending ${\cal A}$   of the SH is only defined after averaging 
 over,
 say, one turn of SH.
At the linker scale, twist-bend and stretch-bend coupling terms are present;
they vanish on the average.
The averaging keeps only the large scale,
rotationally invariant, ``EWLR-like'' contribution.
The remaining terms describe local contributions 
to the bending energy of the SH, 
cancelling each other, hence with no
resulting effect at the fiber scale. The average over a number of linkers sufficient to
recover  the rotational invariance of the fiber (using ${1\over N}\sum_{j=1}^N\vec{u}_j
\sim (\cos z) \vec{A}$) is conveniently replaced 
(except for the ribbon-like structure for which the rotational 
invariance breaks down, see Section V-E)
in the computation by an average over the directions of
$\vec{M}_b$ (denoted $<>$): it has the same effect of extracting only the resulting contribution
at large scale.
After integration of the densities $\epsilon_{bend}$
and $\epsilon_{twist}$ along the linker,
 the elastic energies $E_{bend}$ 
and $E_{twist}$ stored in a linker write:
\begin{eqnarray}
E_{bend}&=&{1\over 2Ak_BT}\;\int_0^l<m_{b}^2(s)>ds\\
&=&{lF^2r^2\over 6Ak_BT}\left(\sin^2\left({\eta\over 2}\right)+
3\cos^2z\cos^2\left({\eta\over 2}\right)\right)+
{lM_t^2\over 2Ak_BT}\sin^2 z-{lFM_tr\over Ak_BT}\cos 
z\sin z\cos\left({\eta\over 2}\right)+
{lM_b^2\over 4Ak_BT}\left(1+\cos^2z\right)
\\
E_{twist}&=&{1\over 2Ck_BT}\;\int_0^l<m_{t}^2>ds={l<m_{t}^2>\over 2Ck_BT}\\
&=&{l\over 2Ck_BT}\left(
Fr\cos\left({\eta\over 2}\right)\sin z+M_t\cos z\right)^2+{lM_b^2\sin^2z\over 4Ck_BT}
\end{eqnarray}

\subsection{Analytic expression of the elastic constants as a function of geometric
parameters of the fiber (SH)}

The quadratic expression of $E_{bend}+E_{twist}$
as a  function of $F$, $M_t$ and $M_b$ 
is straightforwardly identified with (\ref{EWLR2})
 integrated over the length $D$. This justifies
{\it a posteriori} to map the fiber and its
elastic behaviour onto the continuous EWLR model.
In other words, it provides a microscopic validation of the EWLR modeling of the
chromatin fiber. In particular, we check the expected vanishing of the twist-bend and stretch-bend
coupling. Some algebra 
finally yields the elastic constants of
the SH fiber:
\vskip 4mm\hspace{15mm}
\be\label{SHelast}
\mbox{
\begin{tabular}{|rcl|}
\hline
&&\\
\hspace{10mm}${\cal A}$&=&$ \frac{AD/l}{1-{(C-A)\over 2C}\sin^2z} $\\&&\\
\hspace{10mm}${\cal C}$&=&$ {CD\over l}\left({\tan^2(\eta/2)\over 3}+
\cos^2z+{A\over C}\sin^2z
\right)\;\left(
{1\over 1+{\tan^2(\eta/2)\over 3}\left({\cos^2z+{C\over A}\sin^2z\over 3A}\right)}\right)
\hspace{10mm}$
\\&&\\
$\gamma$&=&$k_BT\;{D\over l}\left(\frac{A\cos^2z+C\sin^2z}{r^2\cos^2(\eta/2)}\right)\;
\left(
{1\over 1+{\tan^2(\eta/2)\over 3}\left({\cos^2z+{C\over A}\sin^2z\over 3A}\right)}\right)
 $\\&&\\
$g$&=&${D\over l}\left(\frac{(C-A)\sin z\cos z}{r\cos(\eta/2)}\right)\;
\left(
{1\over 1+{\tan^2(\eta/2)\over 3}\left({\cos^2z+{C\over A}\sin^2z\over 3A}\right)}\right)
$\\
&&\\
\hline
\end{tabular}
}\ee

\vskip 4mm
We underline that our approach yields the elastic constants of any relaxed geometry,
hence allows an analysis of sensitivity with respect to $l$
(i.e. $\tau^0$) and $\Phi$.

\section{Results and physical discussion}

\subsection{Numerical results for the elastic constants}
Numerical implementation of the above analytical formulas
can be performed for any values
of the microscopic parameters $l$ and $\Phi$, in continuation of Section III, thanks to a Maple program.
We here present the results obtained in the two situations
$\Phi=90^o$ and $\Phi=50^0$, taking as a variable
the  linker length.
\vskip 3mm
Starting from the microscopic 
structural parameters $\tau^0$ and $\Phi$, passing through the determination
of the SH geometric characteristics $D$, $r$, $z$ and $\eta$, we obtain
 the explicit values of the elastic constants,
as functions of $\tau^0$ (or equivalently $n_{bp}$) at fixed (arbitrary) $\Phi$.
Indeed, expressions (\ref{SHelast}) are valid for 
any relaxed structure hence  allow 
to analyze the complete range of variations of the elastic constants as $\tau^0$ and
$\Phi$ vary, as shown in Figure 11.
  We here present two typical cases, currently proposed in chromatin
structural studies: $\Phi=90^o$ then $\Phi=50^o$.
A  decrease of $\Phi$  is presumably induced by an increase of salt concentration, lowering
the mutual repulsion of ingoing and outgoing linkers and strengthening the structural
effect of linker histones.
Experimental observations  of a fiber dressed with linker histones showed that
a value of $\Phi$ around $90^o$ corresponds to a salt concentration of less than 5mM NaCl, whereas
$\Phi$ decreases below $50^o$ above 15 mM NaCl \cite{bednar}\cite{H1-holde}.

\vskip 3mm

 The curves of Figure 11 evidence a strong sensitivity of elastic coefficients
 with respect to the fiber structure, as controlled by $n_{bp}$.
A striking result of our study is the sharp decrease of all elastic
coefficients together with the pitch $P$,  around a critical value $n_{bp, c}(\Phi)$
for which $P$ vanishes.
The critical value $n_{bp, c}(\Phi)$
 depends on the value of $\Phi$ and
more generally on the precise modeling of the chomatosome.
As $P$  vanishes, steric hindrance between nucleosomes precludes  to build a
regular structure with $n_{bp}=n_{bp, c}(\Phi)$, so that this feature might be seen as
irrelevant.
Nevertheless, it is possible to have $n_{bp}=n_{bp, c}(\Phi)$
locally, i.e. over less than one turn of SH; this  flexibility,
although local,  might yet  have dramatic
consequences: it is possible to create a noticeable kink in the chromatine fiber
at this point by applying only very weak stresses.
Such ``critical'' turns appear as defects
where the regular
compact fiber is easy to ``open''.
Moreover, the curves near $n_{bp, c}$ but above the excluded-volume threshold $P_c$ are
still  influenced by their vanishing in $n_{bp, c}$.
The steep variation of the elastic constants in the neighborhood of $n_{bp, c}$
is an actual property, corresponding to a highly sensitive fiber: the geometric
parameter $D$ and the elastic constants increase by a noticeable factor within one or
two base pairs.
Another important result, whose implications are discussed below (Section V-D), is the 
observed change
of chirality of the SH, defined in Section III.D as the sign of $ \cos(\eta/2)$
and exactly correlated with the changes in the sign of the twist-stretch coupling $g$. 
\vskip 3mm
Finally, 
the relative energetic contribution of linker bending and linker twisting
 are shown on Figure 12 in pure cases
where only one of the stresses $F$, $M_t$ or $M_b$  is applied (stretch, twist or bend at the
fiber level). 
We checked that the stretching energy and the twist-stretch coupling energy
of the linker 
are negligible
 (atmost a few percent of the total elastic energy stored in a linker)
due to the high value of $\gamma_{DNA}$ (around 1200 pN)  and $g_{DNA}$
(between 20 and 30) \cite{marko-stretch-1}\cite{coeff-g}.
The figure evidences an interesting feature coming from the 
spatial organization of DNA into a chromatin fiber.
According to the fiber structure (i.e. linker length $l$ and angle $\Phi$)
and to the nature of external stresses (pulling force, torsional or flexural torque)
the linkers react  either by bending, either by twisting,
which might play a biological role by regulating the local DNA structure.

\vskip 3mm
Our computation of ${\cal A}$ is based on its ``energetic'' definition through the
expression (\ref{EWLR1}) of the free energy of the EWLR description of the fiber:
the energetic  cost required for
 bending the SH axis and creating  a uniform curvature $\rho$
is $k_BT{\cal A}\rho^2/2$ per unit length.
It originates from the cost of twist and bend distortions induced at lower scale
on the linkers.
This definition does not (and does not have to)  consider possible steric hindrance
(hard-core interactions).
Indeed excluded-volume constraints do not
contribute to elastic energies hence do not
 modify ${\cal A}$; they only forbid some
deformations.
To determine whether a given bending of the SH is possible, one has

-- first to check whether such a bending is ``geometrically'' allowed, i.e. satisfied
excluded-volume constraints ( a necessary condition is that $\rho R_{ex}<1$, i.e. the
radius of curvature $1/\rho$ should be greater that the excluded-volume radius
$R_{SH}$\cite{pieranski},

-- secondly to check whether enough energy is available,
using the value ${\cal A}$ determined above, notwithstanding the
excluded-volume effects.

\noindent
Note that these criterions are independent (geometric and energetic respectively) and
they can be checked in any order.

\subsection{Comparison with an helical coiling (ordinary spring)}

It is interesting to compare the SH elastic behavior to that of a simple helical
``spring'' (a toroidal coiling of DNA), i.e. the continuous structure characterized by
the same angle $z$ (here between its axis and the tangent to the elementary fiber) and
the same radius $r$.
The helical spring is thus implicitely controlled by the same parameters $l$ and $\Phi$ as
the SH.
The comparison enlights:

-- that the chromatin fiber is actually a spring from a mechanical point
of view, although a special, tunable, one;

-- that the difference between ordinary springs and chromatin lies in the
angle $\eta$ ($\eta=0$ in ordinary springs whereas it is not small
in chromatin), from which originates the tunable character of the
chromatin spring.

The reasoning to compute the elastic constants of an helical
spring is at each step analog to the above one (Section IV).
We obtain the following expression for the average energies per unit length:
\begin{eqnarray}
\epsilon_t&=& 
{\left(
Fr\sin z+M_t\cos z\right)^2\over 2Ck_BT}+{M_b^2\sin^2z\over 4Ck_BT}
\\
\epsilon_b&=&
{\cos^2z\;F^2r^2\over 2Ak_BT}+
{M_t^2\sin^2 z\over 2Ak_BT}\;-\;{FM_tr\cos 
z\sin z\over Ak_BT}+
{M_b^2\left(1+\cos^2z\right)\over 4Ak_BT}
\end{eqnarray}
where the average is performed either on the orientations of $\vec{M}_b$, either on a
turn, i.e. over an arclength $2\pi r/ \sin z$ corresponding to the pitch $2\pi r\cot
z$. Such an average is essential to fit in the definition of elastic constants
(continuous, rotationally invariant EWLR model).
The expression of the elastic constants follows:

\vskip 4mm

\be\mbox{Ordinary spring\hspace{5mm}
\begin{tabular}{|rcl|}
\hline
&&\\
\hspace{10mm}${\cal A}$&=&$ \frac{A\cos z}{1-{(C-A)\over 2C}\sin^2z} $\\
&&\\
${\cal C}$&=&$ {\cos z}\left(
C\cos^2z+A\sin^2z
\right)$\\
&&\\
$\gamma$&=&${k_BT\cos z\over r^2}\left(A\cos^2z+C\sin^2z\right)$ \hspace{10mm}\\
&&\\
$g$&=&${(C-A)\sin z\cos^2 z\over r}$\\
&&\\
\hline
\end{tabular}}\ee
\vskip 4mm
In fact, these classical  formulas 
of spring mechanics can be recovered  directly from those obtained for
the chromatin SH by letting $l\rightarrow 0$ at fixed  $z$ in the SH formulas;
accordingly,  $\eta\rightarrow 0$ and $D/l\rightarrow \cos z$.
This link proves that chromatin is actually a special kind of spring. 
In the chromatin fiber, $\sin(\eta/2)$ is not small
and can even reach 1. Hence all the features depending on $\eta$ will exhibit  a striking
difference when comparing the SH and the helical coiling of DNA.
{\it This is the angle $\eta\neq 0$, reflecting the
discrete character of the chromatin assembly, 
 which is mainly responsible
of the tunable elasticity and tunable chirality of the chromatin fiber},
as can be seen by comparing the elastic constants of the helical spring
and of the chromatin fiber.

Let us also compare  the distribution of DNA elastic
energy between twist and bend degrees of freedom.
The comparison is presented in Figure 12.
When only a torque (either torsional, either flexural) is
applied, the partitions in the SH case and in the case of an helical coiling
are exactly identical.
When only a pulling force is applied, 
the energy is mainly stored in the torsional energy in the case of a simple helical spring
whereas it is the converse in the chromatin SH.
In the chromatin fiber, the energy stored in the linker bending is in any cases larger
(and sometimes far larger) than the energy stored in the twist.
This remark reflects the following property of the SH:
given $m_t$ and $m_b$
(modulus averaged along a linker), it is possible to determine a force $F$ and a torsional torque
$M_t$ creating $m_t$ and $m_b$ on each linker only if $m_b\geq {\rm Cte}. m_t$,
where the constant depends on the SH geometric characteristics.
In particular, it is impossible to have $m_b=0$ unless $m_t=0$, i.e. $F=M_t=0$.

This difference between the elastic behavior of the SH and the corresponding simple
helical  coiling originates from the discrete structure of the SH, where the linkers
are spaced by rigid kinks (the nucleosomes).
This
strongly modifies the relation between the pulling force applied at the fiber ends and the local flexural 
($m_b$) and torsional ($m_t$) components determining the DNA elastic energy.
The discrepancy  is maximal for $\sin(\eta/2)=1$, which corresponds 
to the SH conformation in which the linkers cross the SH axis and SH chirality
changes, which obviously never occurs in an helical coiling 
of bounded radius(either left-handed,
 either right-handed, but never in between if $r$ neither vanishes nor diverges).
Although the SH `` looks like'' an helical coiling (see Figure 3), its response to a pulling
force  differs from that of an helical spring. Indeed, we again underline that its
elastic properties are determined by the characteristics of the broken line formed by
the linkers, which might be far different from an helical coiling (especially when
$\eta$ is close to $\pi$, which corresponds to cross-linked structures).

Elasticity theory of an homogeneous cylindrical rod\cite{love}\cite{smith}
predicts the relation $4{\cal A}k_BT=\gamma R_{SH}^2$.
The predicted value of ${\cal A}$ using this formula
is roughly correct for the helical coiling, but it
overestimates the actual value
of the SH persistence length  by a factor of 2 or more. This discrepancy enlights
the importance of the  complex substructure of the SH, 
leading to a fiber far different from an homogeneously filled
cylinder of radius $R_{SH}$ and even from an helical coiling of DNA.

\subsection{The origin of the sensitivity of the elastic constants
 with respect to the fiber structure}

Although the chromatin SH and an helical coiling of DNA exhibit quantitative differences,
their elastic behaviors share the same key feature (see Figures 11 and 13):
a sharp decrease towards 0 of the elastic constants, together with the pitch. Considering a
simple helical spring yields the -- quite intuitive -- explanation of this feature:
when the slope of the elementary (DNA) fiber  decreases to 0 (DNA path almost orthogonal to
the fiber axis), the number of turns per unit length along the  fiber axis increases (see in
particular the divergence of the compaction ratio 10/$D$(nm) on Figure 9,
outside the figure frame).
This accumulation of turns, each acting as an hinge, easily allows a huge deformation
of the fiber, simply by the addition of  small changes occuring in each turn.
This explanation, involving only an accordion-like behavior, ensures the robustness of the
feature. The flexibility dramatically increases when the number of coils per unit length of
fiber increases, whatever the precise geometry with which the elementary fiber is coiled into
a large-scale one.

The strong correlation between the pitch  and the four elastic constants behaviors when
$n_{bp}$ varies simply reflects in the very expression of the elastic constants:
they all write as $(D/l)\;f(z,r)$, hence are slaved to the vanishing of $D$ (or $P$).
The factor $D/l$ shows that the  extensibility and flexibility of the
SH are  controlled by its degree of compaction.
The more compact the conformation (low value of $P$),
the more flexible is the fiber, easily bent, twisted or stretched.
We claim that the result is typical, i.e. does not depend on the details
of the modeling, but only on  the general feature of the chromatin
fiber assembly, i.e. nucleosomes linked by segments of DNA.
This form $(D/l)\;f(z,r)$ also indicates that {\it the elastic properties
of the fiber are mainly  controlled by the angle $z$ between the linkers and 
the SH axis}. Indeed, $z$
roughly determines the decomposition of local stresses among twist and  bend 
degrees of freedom of linker DNA
(stretch is always negligible).

\subsection{A multi-strand spring with tunable chirality}

We first underline that the chromatin fiber is a spring of tunable
chirality: the sign of the twist-stretch coupling
$g$, associated to the chiral nature of the SH,  changes twice 
within a ``period'' (an $l$-interval
of width 10.6 bp, i.e. the DNA pitch), see  Figure 11; correspondingly,
a change from a right-handed SH to a left-handed SH occurs, as seen 
on Figure~1.
Reverse behaviors are observed in
right-handed and left-handed SH.
Indeed, change of chirality exactly corresponds to 
 change in the sign of the susceptibility
$\partial D/\partial \tau$, i.e. the slope of the curve giving $D$ as a function of
$n_{bp}$ (see Figure 8), up to a factor 
$\partial n_{bp}/\partial\tau=n_{bp}^0/2\pi$.
The noticeable consequence is that a change $\Delta\tau$ will either  decondense or compact the fiber according to the sign 
of $\partial D/\partial \tau$, i.e. to the chirality of the SH.

Far more, the two-level structure of the chromatin fiber makes it quite similar to a multi-strand
spring:
an interesting behavior arises from the interplay between the right-handedness of the
linker DNA and the tunable chirality of its coiling within the SH.
Applying (only) a pulling force $F\vec{A}$ creates a torsional torque:
\be 
m_t=Fr\sin z\cos(\eta/2)\ee
The sign of $\cos(\eta/2)$ thus determines the sign of
 $m_t$ (recall that
by definition $\sin z>0$). According to this sign, i.e. to the SH chirality, 
pulling the chromatin fiber will either twist, either unwind the linker DNA double
helix. This behavior is well-known in the context of multi-strand
 springs\cite{costello}.
Its biological interest might be to regulate DNA denaturation required for 
transcription and replication.
We get a similar result when applying a torsional torque  $M_t\vec{A}$;
the torsional torque at the linker level writes $m_t=M_t\cos z$. A change of sign in
$\cos z$ inverts the action of the given torque $M_t\vec{A}$ at the DNA level,
and we see that such a sign alternation actually occurs in each interval of length 10.6 bp.

Conversely, a modification of the linker twist, as can be achieved by intercalator enzymes,
will modify in a tunable way the geometry of the SH fiber;
according to the SH chirality, the same change of the linker twist will
either condense ($\Delta D<0$), either decondense  ($\Delta D>0$)
the chromatin fiber. This possible mechanism 
of condensation/decondensation and its biological relevance
will be investigated in a following paper (see also Section V.F).

\subsection{Special geometries}
\noindent\underline{ Ribbon (or zigzag):}
a first peculiar geometry of the fiber is the zigzag  or ribbon-like structure, 
for which the cross-linked  and roughly toroidal  coiling degenerates into a flat structure.
The number of nucleosomes per turn then reaches its minimal value of 2, i.e. $\theta=\pi$.
In this case, represented on Figure 14, 
\be
\beta={\pi\over 2}-\alpha\hspace{15mm}
\cos z=\sin\beta\sin(\Phi/2)\hspace{15mm}D= {l\sin(\Phi/2)\over\sin\beta}
\ee
We locate this kind of configurations by determining on Figure 9  the linker lengths
for which $2\pi/\theta$ equals 2.  
It appears (see Figure 10) that $\cos z$ 
then  reaches is maximal value.
Note that a ribbon is chiral, and is not the maximally extended configuration;
on the contrary, a small length change ($\Delta n_{bp}=1$) drives it into sterically forbidden
configurations.

\vskip 10mm
\noindent\underline{Nonchiral configurations:}
the SH chirality changes when $\cos(\eta/2)$ vanishes, i.e. $|\eta|=\pi$; in this case, the
linkers cross the SH axis.  This configuration is  remarkable mainly
for its mechanical properties. Indeed,
the absence of chirality amounts, from its very definition,  
to the vanishing of the twist-stretch coupling: $g=0$. Moreover, as explained in
Section V.D, 
 the sign  of $\cos(\eta/2)$
determines whether the linker DNA is unwound (i.e.  denatured)
 or on the contrary twisted when the SH is pulled. 

Since $\sin(\eta/2)=1$, the configuration is the  limiting case 
opposite to the helical spring, for  which $\sin(\eta/2)=0$.
In particular, when the fiber is only pulled
($\vec{M}=0$) the twist energy stored in the linker vanishes: all the elastic energy of
the linker is stored in the bending degree of freedom:
\be
E_t=0\hspace{15mm}E_b={lF^2r^2\over 6Ak_BT}={l^3F^2\cos^2(\Phi/2)\over 24\, Ak_BT}>0
\ee
which never occurs in an helical spring.
We underline that non-chiral structures  (special linker positioning) and ribbon-like
structure (special nucleosome positioning) should not be confused.

In fact,  
this configuration 
is not rotationally symmetric, hence has 
two bend elastic  constants. They are computed within the same 
lines as above, but without performing any average over the direction
of the bending torque $\vec{M}_b$; the identification is done with an
asymmetric EWLR ($4\times 4$ matrix with two bend persistence lengths).
This yields:
\be
{\cal A}_1={DA\over l}  \hspace{10mm} {\rm and}{\cal A}_2={DA\over l}
\;{1\over 1-{(C-A)\over C}\sin^2z}
\ee
 Nevertheless, the orientation 
of the bending torque is not controlled at this level, hence
the relevant elastic behavior, to be observed experimentally,
is rather that predicted after having performed an average over
the torque directions. This recovers a rotationally  symmetric behavior, 
characterized
by the  above given persistence length ${\cal A}$, as it can be seen directly
in the formula:
\be
{1\over {\cal A}}={1\over 2}\;\left({1\over {\cal A}_1}+{1\over {\cal A}_2}\right)
\ee

\vskip 5mm
\noindent\underline{Columnar packing:}
special configurations are obtained when $n_{bp}$ is close to an integral multiple of
$n_{bp}^0$ (for instance $n_{bp}=42$ bp or 43 bp).
In this case, nucleosome axes are all parallel 
hence parallel to the SH axis due to the rotational symmetry.
We actually check on the curves of Figure 10 that $\beta=0$ for $n_{bp}= 42$ bp,
 whatever $\Phi$.
It is noticeable that  geometric characteristics as $D$, $z$ and $\eta$ are much
sensitive to variations  of $n_{bp}$ around this value of 42 bp,
which reflects in a similar sensitivity of the elastic constants, as seen on Figure 11.
Fine tuning of $n_{bp}$ and $\Phi$ leads to a configuration in which the nucleosomes are
organized in columnar arrays.
It has been evidenced\cite{livolant1}\cite{livolant2} that free nucleosomes
 exhibit a liquid-crystal-like
nature and tend to form columnar phases, in which the nucleosomes are stacked one upon
each other with  their axes parallel to  the column.
A SH configuration exhibiting a columnar packing of nucleosomes in low-salt conditions
is likely to be stabilized by internucleosomal interactions when the ionic force
increases, better than any other configuration. In presence of linker histones, the
specificity of ``columnar'' SH is strengthened since linker histones too exhibit a
propensity to stacking.

\subsection{Comparison with chromatin fiber pulling experiments}
We carry on the discussion by presenting the experimental validation and prospects of
our results.

In the past few years, a novel experimental method of investigating structural and
mechanical properties  of biological macromolecules or complexes, as DNA or
chromosomes, has been
developped\cite{smith}\cite{croquette}\cite{libchaber}\cite{poirier}.
It is based on micromanipulations and force measurements on an isolated fiber.
It belongs to the rapidly expanding  field of investigations known as ``single
molecule biophysics'' \cite{singlescience}.
The experiment deviced in the context of chromatin study transposes to a chromatin
fiber a methodology first implemented with DNA. It consists in pulling the fiber under
various constraints and in various conditions (salt concentration, presence of specific
enzymes or chemical factors), and to determine the ensuing deformations of the fiber. 
Varying  the pulling force yields force-extension curves characterizing the elastic
response of the fiber. 

Nevertheless, at the chromatin scale, the applied stresses are artificial 
and cannot sensibly refer to an event occuring {\it in vivo}:
contrary to DNA case, an enzyme is not  large enough to directly handle the 30-nm fiber
or to experience the chromatin fiber state of strain, for example its torsional
strain or its curvature (the only {\it in vivo} mechanism of that kind, i.e. the
action of the mitotic spindle on chromosomes\cite{duplantier}, occurs at a much larger
scale).
The biochemical processes, for instance binding of a biological factor or chemical
modification, occur at the elementary level (scale of DNA and nucleosomes), 
whereas the experiment probes the 30-nm fiber behavior.
A mechanical modeling relating the DNA scale and the fiber scale is thus necessary  to
exploit all the informations provided by single-fiber pulling experiments
in terms of biological functions.
Conversely, these experiments are essential to validate  the model and the underlying
hypotheses on structure and interactions, to fit dubious parameters and possibly to
ask for refinements.
The EWLR model allows to predict force-extension curves, to be compared with those obtained
in single-fiber experiments.
Our study, describing the microscopic
root of the EWLR model of chromatin,  thus provides a bridge between microscopic structure
and observable properties of the overall fiber.

In the experimental setup used by Cui and Bustamante \cite{cui-bust},
the fiber, of relaxed length $L$, is pulled by means of optical tweezers. 
There is no direct contact with the
fiber ends, which are free to rotate. No torque is applied ($\vec{M}=0$)
and free twist fluctuations take place.
It corresponds to the situation recalled just above.
Some of  force-extension curves that they present
are relaxation curves in low salt, for which it is legitimate to ignore interactions between
nucleosomes. 
Cui and Bustamante fitted these experimental 
relaxation curves within the EWLR model
presented in Section IV.A. Recalling
that their experimental setup allows free twist fluctuations, the fit involved only
two elastic constants: the bend persistence length ${\cal A}$ and an effective stretch
modulus $\gamma_{eff}=\gamma -k_BTg^2/{\cal C}$. They found 
${\cal A}\approx 30$ nm and $\gamma_{eff} \approx 5$ pN, values to which we may compare our
theoretical predictions.
\vskip 3mm

The comparison leads to a striking result: we evidence on Figure 11 that the set of
values (${\cal A}\approx 30$ nm, $\gamma_{eff}\approx5$ pN) is obtained 
  for $n_{bp}$ belonging to a narrow window between 42 and 43 bp whatever the value of
$\Phi$ is (we checked values of $\Phi$ ranging from 130$^o$ to 40$^o$).
Since $n_{bp}$ is then an integral multiple of
$n_{bp}^0$, this case corresponds to special structures of the SH in which the nucleosomes pack into
columnar arrays (IV.E); only the number of columns varies with $\Phi$ and with the precise
modeling of the chromatosome, see Figure 3.
Pulling the fiber will induce a change in $\Phi$ but not in $\tau$, due to the special
orientation of the nucleosomes;
this orientation is thus preserved,  hence $\gamma$ and ${\cal A}$ will not change: no
nonlinear effect arises when starting in this special
columnar  geometry, hence a fit by an EWLR
actually accounts for the whole force-extension curve, even in the region where $u$ is not
small with respect to 1.

\vskip 3mm
In an experiment performed at higher salt concentration (in 40-150 mM NaCl)
Cui and Bustamante
observed a plateau in the
force-extension curve at a value $F_c$ between 5 pN and 6  pN.
They interpret this plateau as a structural transition corresponding to the 
breaking of some short-range 
attractive interactions between chromatosomes, from which follows a
dramatic decondensation of the fiber, at constant force.
Our modeling cannot, of course, directly account for this conformational transition since
internucleosomal interactions are ignored: only relaxation curves can
be predicted, or curves in low salt conditions where interactions are always
negligible, even at low force, and the chromatin always extended (there is besides no plateau in this
low-salt case).

But we claim that our study is specially well-suited to investigate 
the relevant biological question, which  is rather  
 to determine local mechanisms that could induce
such a decondensation, i.e. that could create local stresses
 of large enough strength
to break the interactions.
Indeed, a simple criterion of efficiency of biological factors 
(e.g. enzymes) is that the global
stress resulting from their binding should be  larger
 than the critical force $F_c$ measured
by Cui and Bustamante.
Our approach provides a direct method  to check this criterion since it
 relates the
local strains  induced by the binding of the enzyme
on linker DNA (experimentally measured or deduced from
a molecular mechanics simulation) 
first to the  associated local stresses
(equation \ref{linearDNA});  then, 
by extending the computations presented in Section IV, 
it is  possible to relate these local
stresses to the global stress experienced by the chromatin fiber, 
 to be compared to the measured value $F_c$, and to the global deformation
of the fiber conformation.
The implementation of  this methodology in the case of intercalators,
which
  gives insights on the 
 {\it in vivo} condensation/decondensation mechanisms of the chromatin fiber,
will be presented in
a subsequent paper.

\vskip 3mm
We mention that our work is not the first theoretical, model-based approach aiming 
at accounting for these experimental results.

A first one has been developped by Katritch et al.\cite{simul-bust}, using a Monte Carlo
simulation of the fiber.
One of the interests of such a simulation is to include explicitly thermal fluctuations but,
as they show in their paper, the behavior experimentally observed 
is mainly deterministic (except at very low force).
Moreover, the  possibility offered by a simulation to treat accurately the interactions, for
instance between nucleosomes, is at the moment hampred by the lack of data, preventing to go
beyond an effective isotropic  model of interaction.
We thus believe that Monte Carlo simulations are not necessary, at least 
as concerns  the problem of
reproducing the experimentally observed force-extension curves.
Indeed, an analytic answer is at hand, even with nucleosome interactions,
and with more refined excluded-volume constraints (see Figure 7 and the analog that could be
constructed for the linkers, by unwrapping the cylinder of radius $r$).
Simulations would become really useful to handle a more detailed description of the
chromatosome, for instance with an explicit linker histone.

A second approach has been performed recently by Schiessel et al.\cite{schiessel}.
As discussed below (VII.B), its basic step is to relate the strains at the DNA and fiber
scales (instead of the stresses, as performed here in Section IV). They only managed to
determine some elastic constants 
(namely $\gamma$ and ${\cal A}$) of special geometries,
and at the expense of some approximations,  but
they give analytical expressions directly in terms
of the microscopic parameters ( here $l$ and $\Phi$). Moreover, 
the 
forces that they predict in their theoretical force-extension curves
 are smaller  
 than the results of Cui and Bustamante by a noticeable factor of 4. 
This could be explained by  their derivation, summing up 
the bend  and twist contributions to
$\gamma^{-1}$, which is questionable. Indeed, looking at the basic example 
of an helical spring shows that
instead of writing   $\gamma^{-1}(A,C)=\gamma^{-1}(A, C=0)+\gamma^{-1}(A=0, C)$,
one should sum up the stretch moduli, according to
  $\gamma(A,C)=\gamma(A, C=0)+\gamma(A=0, C)$
as in the case when springs act in parallel. 
 This leads to a
discrepancy by a factor of 4 when the ``elementary'' strech modulus
$\gamma(A, C=0)$ and $\gamma(A=0, C)$ are of the same order.
Moreover, in the case of the SH, the actual
 expression -- see equation (\ref{SHelast})-- shows 
that such an additive
decomposition does not exactly hold in the SH case.
 These two objections  could explain the discrepancy between  
 our value of 5 pN for $\gamma$, in agreement  with the experimental
results of Cui and Bustamante, and  their value $\gamma=1.2$ pN,
yet observed for quite similar fiber conformations, with crossed linkers
(Schiessel et al. considered a structure with (in our notations)
$\Phi=85^o$ and $\tau=36^o$, which corresponds to  42-43 bp in our model).


\section{Biological discussion}

\subsection{A tunable, highly sensitive,  elastic structure}

We thus evidence a wide range of different elastic behaviors, separated by a minor
change in the linker length. We suggest that this tunable elasticity might be used as
a regulatory mechanism during the cell cycle.
For instance, a slow modulation of the linker lengths,
 might create different domains
in the chromatin fiber, of much different rigidities, and might  provide a preliminary 
underlining of transcriptionally active chromatin regions.
The response of these different regions to a same local stress (protein binding, for
instance) will be dramatically different. For example, intercalation might condense or
decondense the fiber, according to the sign of $\partial D/\partial\tau$,
i.e. to the chirality:
within $\Delta n_{bp}=2$ bp, opposite consequences will be observed.
This underlining, inscribed in the very structure of the fiber,   allows a rapid
and selected response to a non-specific stress, which might be 
 biologically more relevant
than a mechanism based on enzyme recognition of a specific sequence,  moreover
possibly buried inside the
fiber. Mechanical sensitivity is likely to provide efficient switches for processes
occuring at the fiber level. 
\vskip 10mm
The chromatin fiber thus exhibits  tunable structure, tunable
chirality and tunable
elastic
properties.
 We suggest three possible  mechanisms to implement 
the required adaptation of the linker length: 

-- the first mechanism involves nucleosome displacement; acetylation of histone tails
untightens the DNA wrapping around the histone core and 
presumably allows nucleosome mobility. Nevertheless, topological (linking number
conservation) and mechanical (helical gearing) constraints make the motion of the
nucleosome far different from a mere translation along the DNA,
and the kinematic feasibility deserves to be investigated further;

-- the second mechanism involves intercalating enzymes, modifying the twist of the
linker. A detailed study of the interplay between linker intercalation and the
chromatin fiber mechanics will be presented in a subsequent paper;

-- the best candidate might be a mechanism involving the linker histone. Indeed,
the value $n_{bp}$ involved here is the  effective length of the linker, beyond linker
histone (i.e. outside the chromatosome). This length might be tuned by a slight displacement of the linker histone
away ($\Delta n_{bp}<0$) or towards  ($\Delta n_{bp}>0$) the nucleosome.
\vskip 1mm
\noindent
An alternative tuning mechanism lays on the variation of $\Phi$, controlled in particular
by the
presence of linker histone,  salt concentration and histone tails binding affinities.

\subsection{A novel chromatin structure}

We evidenced that
a value of linker length $n_{bp}$ between 42 and 43 bp leads to the
  values ${\cal A}\approx 30$
nm and $\gamma\approx 5$ pN whatever $\Phi$ .
Since $n_{bp}$ is then about
$4\;n_{bp}^0$,  the nucleosome axes are all parallel
hence parallel to the SH axis due to the rotational symmetry.
That such organized structures lead to ${\cal A}=30$ nm and $\gamma_{eff}=5$ pN,
whatever $\Phi$ is (and even whatever $r_{nucl}$ and $H$ are, as we checked)
is explained by the fact that the SH elastic properties
originate from the linker DNA contribution.
As shown by the computation of SH elastic constants, this contribution is mainly fixed
by the linker orientation with respect to the SH axis (i.e. $z$).
This matching between our predictions and experimental results strongly suggests that
the structure underlying the observed elastic properties is a 
columnar packing of nucleosomes.
We thus supplement  the argument of Yao et al.\cite{yao} in favour to
rotationally phased
nucleosomes: the actual relaxed SH structures are selected according to their ability
to be stabilized by internucleosomal interactions
when the ionic force increases. Our claim is supported by the results  of
Livolant et al.\cite{livolant1}\cite{livolant2}, in which it is observed that
 nucleosomes exhibit a liquid-crystal-like nature,
leading to a spontaneous columnar ordering.
We thus expect native SH structures to favour columnar packing of nucleosomes, as they
correspond to the more  robust three-dimensional organization of the chromatin
fiber. Note that whereas  a model with straight linkers is
compatible with the observed structure and elasticity of the chromatin fiber
at low ionic strength (5 mM NaCl), interactions between 
stacked nucleosomes should induce a
bending of the linkers
at higher ionic strength. More probably, the conflicting effects
consistency between 
nucleosomes stacking and linker stiffness might
 be reconciled by linker DNA kinks,
occuring near the entry/exit points and induced by the 
binding oh linker histones H1 . 

For $\Phi=90^o$ (low-salt situation), the degree of compaction $10/D$(nm)
reaches its minimal value for the same value $n_{bp}$ 
located between 42 and 43 bp. This underlines a key
feature of the corresponding  configuration: at low salt, it is the most extended and
rigid configuration; at the same time, it is the most responsive to salt induced
compaction.
Indeed, as seen on Figure 9,  the degree of compaction $10/D$(nm) is minimal at $n_{bp}=42$ or 43 bp for
$\Phi=90^0$, and it strongly increases when $\Phi$ decreases to $50^0$, which is 
an acknowledged effect of increasing the salt concentration. 
Moreover, this configuration strongly favours a second compaction step, ensured
by the
attractive  interactions between nucleosomes (or rather chromatosomes)
that arise when the nuclesome faces are close enough.
We in particular recover in this scheme the two-stage compaction of the 30-nm chromatin
fiber observed experimentally\cite{patrone1}\cite{patrone2}.
An insight on this interaction-induced compaction can be obtained by setting the
effective parameters $r_{nucl}$ and $H$ to 0, thus mimicking the enhanced influence of
linker histone  at high salt; in this case, $P$ decreases to about 6 nm, indicating that
nucleosomes actually stack very closely onto each other and lead to a superstable (and presumably
rigid) fiber. It has been suggested\cite{H1-holde} that H1 is required
 not so much to get a folded
fiber (compact fibers have been observed in absence of H1) but to get a properly
folded fiber.
We suggest that H1 might be involved in the tuning of the effective linker length and
twist, actually involved in the assembly. In any cases, H1 stacking interactions 
favour configurations exhibiting  columnar arrays of nucleosomes.

\section{Conclusion and perspectives}

\subsection{What did we learn from this study?}




The chromatin
sensitivity of the fiber response to global or local stresses
evidenced in this study sheds light on the
biological interest of the so peculiar and so universal assembly of chromatin fiber.
It enlights possible
relations between small-scale structure and gene regulation
through the
fiber mechanical properties.
For instance, we evidenced that a minor change of $l$ or $\tau$ around some ``critical'' value  
inverts its chirality. 
Hence, its response to a torsional torque  induces either
an extension, either a contraction of the fiber, and its response to a pulling force
will either winds or unwinds the SH fiber.
Moreover, as DNA itself is chiral (right-handed), the response at the DNA level is
also controlled by the SH chirality: pulling the fiber unwinds the DNA if the SH is
left-handed.

\vskip 3mm
The tunable energy partition between twist and bend degrees of freedom at the
linker DNA level may be of  biological interest: according to
the chromatin configuration, either a twist-sensitive protein 
will bind onto linker DNA, either a protein whose binding 
is favoured by the local curvature of linker DNA  will be the
adapted factor. Hence the mechanical sensitivity can participate
to  biological recognition or
specificity. 
Conversely, we read on the associated  curves (see Figure 11)
 whether  a twist-modifying protein
 (intercalator, gyrase), for instance, may induce a required strain of the SH.

\vskip 3mm
 Comparison with experimental results give clues about
 the much debated chromatin structure\cite{revue}. It brings about a novel structure:
a columnar packing of nucleosomes.
We suggest that a possible role of linker histone might be  to select the proper structure, at low salt,
by tuning the linker length so as to have $n_{bp}/n_{bp}^0$ equal to some integer.
Also, a tuning of $\Delta n_{bp}= 1$
or 2 bp might be achieved by enzyme intercalation.
When ionic strength increases, compaction takes place, first due to a decrease of
$\Phi$ (which keeps the nucleosome orientations unchanged, along the SH axis); then,
when nucleosomes happen to be stacked in columns, the compaction is ensured by
the interactions between stacked nucleosomes, between stacked linker histones
(counterbalancing the linker DNA repulsion) or between histone tails and nearby linkers
or nucleosomes.


\subsection{A general method for studying elasticity of linear complex fibers}

 The problem of relating the elastic coefficients of the 
chromatin fiber  to the geometric and elastic
properties of the underlying ``microscopic'' structure
(assembly of nucleosomes and linkers) is reminiscent of similar  works
performed for DNA by Marko and Siggia\cite{markosiggia}
and 
O'Hern et al. \cite{kamien}\cite{coeff-g}.
They described the dsDNA at two levels: as an helical coiling at small scale and as an
EWLR at a slightly larger scale.
They derived similar formulas relating the elastic
coefficients ($C$, $A$, $\gamma_{DNA}$ and $g_{DNA}$) to the
geometric and  elastic parameters of the underlying helical model.

 We point that their computation rests on the relation between the
``microscopic''  strains and the dsDNA strains, plugged into the equality of the EWLR
free energy and the free energy computed within the microscopic model.
Generally, numerous sets of microscopic strains
achieve the macroscopic strains,  but only the set of lowest free energy yields the
actual free energy of the EWLR.
Relating properly microscopic and macroscopic strains
thus requires to minimize the small-scale free energy, given the macroscopic strains.
Performing this minimzation is nothing but writing the conditions for
the local equilibrium of the assembly.

When the microscopic model is homogeneous, the conditions for local equilibrium simply
expresses in the extensivity of the strains. The uniformity of the local
strain  densities  thus allows to relate them to global strains without an explicit
minimization (think to identical springs in series).
This works for DNA\cite{coeff-g} but not in the case of chromatin.
The discrete and complex nature of the chromatin assembly
leads to difficult and cumbersome computations in order to determine the linker shape,
as it can be seen in the work of Schiessel et al.\cite{schiessel}.
Moreover, this approach fails to give an analytical solution except when the relaxed
fiber exhibits a special geometry, for example a ribbon-like  flat structure.

Determining the conditions for local equilibrium is precisely what is done, more
straighforwardly, in our approach. We indeed write equilibrium equations given 
the global stresses and solve them to get the local stresses arising in each point of
the assembly (at equilibrium under the given global stresses).
Relating the stresses  at the microscopic and at the fiber levels
follows from the basic principles of classical mechanics;
this method  appears to be at the same time more
simple and more easily generalized. It is in fact the only way to bridge the 
linker elasticity to the SH elasticity in any geometry.
Moreover, it extends to more complex situations as intercalated linker DNA
or more generally, situations where forces and torques are applied at the DNA level.

We thus underline that the proper method to express the elastic properties of an
assembly as a function of the elastic properties of the basic elements 
is to relate the global stresses, applied to the assembly, and the local stresses
experienced by the lower scale elements.
This relation can be used in both ways, to investigate

--  either small-scale 
repercussions of a global stress, hence how a global stress
(as those applied in micromanipulations) can be used to probe the
fiber at the elementary level,

-- either large-scale response to local stresses, hence how biological factors binding
on the linker DNA could induce major structural and conformational changes in the
overall fiber.

\subsection{Biological perspectives}

Our study underlines that the mechanical properties
of special structures, selected according to the phasing of the nucleosomes therein, 
might be involved as a regulatory factor in the chromatin biological function.

Having modelled the 30-nm fiber as a EWLR, with explicit values of the elastic
coefficients, a natural extension of our study is 
 to consider higher levels of organization:
a plectonemic coiling, leading to a 60-nm fiber,
or an helical coiling, whose elastic properties follow from classical spring
mechanics. 
The question is then to unravel  the implications of the chromatin structure and 
its elastic
properties on the higher levels of organization.
A mechanical approach similar to that implemented in this paper
is essential  to bridge electron-microscopy structural observations evidencing
fiber-like objects at higher levels (60-nm fiber, ``chromonema'' fiber of
diameter 100-130 nm\cite{belmont1} and chromosome) and 
experimental results on  chromosome elasticity
 obtained by pulling a single chromosome\cite{libchaber}\cite{poirier}.
The challenge is to understand  the mechanics of the chromosome and its involvment
in the biological functioning of the chromosome throughout the cell cycle.

\vskip 3mm
Another direction in which to exploit the results of the present paper
is to determine the stresses that can be exerted at
the fiber scale by groove-binding proteins or intercalators
 when they bind onto linker DNA.
For instance investigate whether a local  decondensation of the fiber might be induced by intercalating enzymes
and controlled by linker lengths.
More generally, our approach is a privileged tool to investigate the 
action at the fiber level of small-scale biochemical stresses (protein binding,
histone tail acetylation), then to
 describe how they can act as mechanical
switches  and exploit the tunable elasticity of the fiber
into regulatory schemes.

\newpage
\section{Captions}

\vskip 5mm\noindent
{\bf  Figure 1:}
 Geometric description of linker DNA (section II.A).
{\it Above:} the arrows indicate the orientation
of the two strands, delimiting the minor groove (filled in grey) and the major groove.
The vector $\vec{t}(s)$ is tangent to the minor groove and rotates around the local axis
$\vec{u}$ (straight in a relaxed linker) with an angular rate
$\omega^0=2\pi/l^0$; the two vectors make a constant angle $90^o-\alpha_{DNA}\approx
62^o$. {\it Below:} tranverse view; the  bold contour locates the minor groove.

\vskip 5mm\noindent
{\bf  Figure 2:} Nucleosome modeling and assembly of the chromatin fiber
(section II.B).
{\it Left above:} view in perspective of the nucleosome $j$; $\alpha$ is the slope of
the DNA left-handed wrapping around the histone core
($\alpha\approx 4.47^o$). The dyad axis 
$\vec{D}_j$ is orthogonal to the
nucleosome axis $\vec{N}_j$. {\it Left below:} view from above (projection on a plane
orthogonal to the
nucleosome axis. {\it Right:} vectors $\vec{N}_j$, $\vec{u}_j$ and $\vec{t}_j(l)$
in the plane tangent to the nucleosome at $E_j$.

\vskip 5mm\noindent
{\bf  Figure 3a:}
Various SH structures obtained when the parameter $l$ (equivalently
$n_{bp}$) is varied, for $\Phi=90^o$.
 Note the columnar packing ($\beta\approx 0$) obtained for $n_{bp}$ 
between 42 bp and 43 bp, 
and the ribbon-like structure (two nucleosomes per turn, $\theta=\pi$)
obtained for $n_{bp}=47$ bp, 
whatever the value of $\Phi$.
In between , the fiber exhibits cross-linked configurations.
In each period of length 10.6 bp, an interval (depending of $\Phi$)
 of $n_{bp}$-values
is forbidden  as it corresponds to self-overlapping configurations
(here around  $n_{bp}=38$ bp)

\vskip 1mm\noindent
{\bf  Figure 3b:}
Various SH structures obtained when the parameter $l$ (equivalently
$n_{bp}$) is varied, for $\Phi=50^o$.
Configurations around $n_{bp}=45$ bp are forbidden (steric hindrance).

\vskip 5mm\noindent
{\bf  Figure 4:} {\it Left:} SH from above (projection in a plane 
orthogonal to the SH axis $\vec{A}$,
here for $\beta=90^o$.  
We define $R$ as the  distance  from any nucleosome center $G_j$ to the SH axis
(measured along the dyad axis $\vec{D}_j$), $r$ as the  
distance  of any entry (or exit) point
 to the SH axis $\vec{A}$; 
the ``excluded-volume'' radius $R_{SH}$ of the SH, 
i.e. the radius of the cylinder
of axis $\vec{A}$ containing the whole fiber (including the nucleosomes)
satisfies $R_{SH}=R+r_{nucl}$ due to the peculiar orientation of the nucleosomes
(the axes $\vec{N}_j$ are tangent to the cylinder of axis $\vec{A}$ and radius $R$
whatever $\beta$). 
We also introduce the angle $\theta$
of the rotation 
around the SH axis $\vec{A}$, transforming the projection of a nucleosome
into the following one, and
the angle $\eta$
between the projections 
of the  vectors relating $S_{j-1}$ and $E_j$ to
the SH axis.
 This angle  $\eta$ is chosen in $[0,2\pi[$. It satisfies $l\sin z=2r\sin(\eta/2)$.
{\it Right:} projection of the first nucleosome in the plane
orthogonal to its axis $\vec{N}_1$.
Note that the nucleosome dyad axis is orthogonal both to the nucleosome axis 
(nucleosome symmetry) and to
the SH axis (symmetry of the assembly).

\vskip 5mm\noindent
{\bf  Figure 5:} {\it Left:}
Location of the nucleosome axis $\vec{N}_j$ in the frame spanned by
the SH axis $\vec{A}$ and the nucleosome $j$ dyad axis $\vec{D}_j$;
it lays in the plane orthogonal to $\vec{D}_j$ and makes an angle $\beta\in[0,2\pi]$
with respect to $\vec{A}$. This angle, varying with $\Phi$ and $l$,
 satisfies $\cos\beta=\vec{N}_j.\vec{A}$.
{\it Right:} Location of the linker $\vec{u}_j$
 in the frame spanned by
the SH axis $\vec{A}$ and the nucleosome dyad axis $\vec{D}_j$;
the linker  makes a constant angle $\xi$ with the dyad axis.

\vskip 5mm\noindent
{\bf  Figure 6:} 
Front view of the SH projected  in a plane spanned by the SH axis $\vec{A}$ and the linker axis
$\vec{u}_j$. The  SH axis is actually ahead of the linker, which lays in the figure plane;
$E_{j-1}$, $G_{j-1}$, $S_j$ and $G_j$ are projections on the figure
plane. 
Only the track of the nucleosomes is indicated: neither their axes  nor their dyad axes
belong to the drawing plane.
$D$ is the 
distance  between two successive  nucleosomes along the SH axis $\vec{A}$,
whose orientation  is chosen so that $D>0$.
We denote $z$ the angle  between the linker and the SH axis;
it satisfies $\vec{u}_j.\vec{A}=\cos z$ and 
by convention 
$z\in[0,\pi[$ hence $\sin z>0$.

\vskip 5mm\noindent
{\bf  Figure 7:} Unwrapping of the cylinder of radius $R$
(containing the nucleosome centers). 
The construction only requires to know
the values of  $\beta$, $\theta$, $D$
and the nucleosome dimensions $r_{nucl}$ and $H$. This drawing can be conveniently used to check
the excluded-volume constraint between any two nucleosomes.
Linkers, crossing the plane of the drawing, are not represented, nor the entry and
exit point (located outside the drawing plane).

\vskip 5mm\noindent
{\bf  Figure 8:}
 Fiber actual diameter $2R_{SH}$ ({\it dashed upper line, curve (1)}) 
and pitch $P$ ({\it solid line, curve (2)}) in nm, versus linker length (in bp), for $\Phi=90^o$ ({\it left})
and $\Phi=50^o$ ({\it right}).
As what matters is the twist angle $\tau=2\pi n_{bp}/n_{bp}^0$, the number $n_{bp}$
of bp is allowed to vary continuously.
The thin line {\it (3)} represents the distance $D$ between two nucleosomes along the SH axis.
The  line made of crosses {\it (4)}  represents the lower bound $P_c$ on the pitch
(equation \ref{pitchbound}). The configurations with $P>P_c$ satisfy excluded-volume
constraints.

\vskip 5mm\noindent
{\bf  Figure 9:} ({\it Solid line})
compaction rate $10/D$(nm) (number of nucleosomes per 10 nm of fiber)
and ({\it crosses}) number $2\pi/\theta$ of nucleosomes per turn
versus linker length (in bp), for $\Phi=90^o$ ({\it left})
and $\Phi=50^o$ ({\it right}).
Two nucleosomes per turn correspond to ribbon-like configurations, 
whereas the maximal number is reached for columnar packing ($n_{bp}=43$ bp).

\vskip 5mm\noindent
{\bf  Figure 10:} Cos$(\eta/2)$ ({\it solid line, (1)}), $\cos\beta$
 ({\it dashed line (2)}) 
 and $\cos z$ ({\it crosses (3)}) 
versus linker length (in bp), for $\Phi=90^o$ ({\it left})
and $\Phi=50^o$ ({\it right}).
Cos$(\eta/2)=0$ corresponds to a change in the SH chirality; $\cos\beta=1$
corresponds to columnar packing in which the nucleosomes axes are parallel to the SH
axis: $\cos z$ is then small and the linkers are almost orthogonal to the axis
(the difference is $\alpha\approx 4.5^o$ where $\tan\alpha$ is the
slope of the linkers on the nucleosomes).

\vskip 5mm\noindent
{\bf  Figure 11:}
{\it Above:}
Twist persistence length ${\cal C}$ ({\it dashed line, curve (1)}),
bend persistence length
${\cal A}$ ({\it solid line, curve (2)})
 and SH pitch $P$ ({\it lower thick line, curve (3)}) in nm
versus linker length (in bp, according to $l=l^0n_{bp}/n_{bp}^0$)
for  $\Phi=90^o$ (left) and   $\Phi=50^o$ (right).
Note that they all vanish at a critical value $n_{bp,c}$, strongly depending on $\Phi$
and on the detailed modeling of the nucleosome (or chromatosome).
The dotted line  ({\it curve (4)}) represents the lower bound $P_c$ (equation \ref{pitchbound})
indicating the values of $n_{bp}$ forbidden by steric hindrance.
{\it Middle } stretch modulus $\gamma$ ({\it solid line, curve (1)})
and effective stretch modulus $\gamma_{eff}=\gamma -k_BTg^2/{\cal C}$
 ({\it dashed line, curve (2)}) in pN
for 
$\Phi=90^o$ (left) and $\Phi=50^o$ (right).
{\it Below:} twist-stretch
coupling $g$ (no dimension) versus linker length (in bp) for 
$\Phi=90^o$ (left) and $\Phi=50^o$ (right).

\vskip 5mm\noindent
{\bf  Figure 12:} these figures represent the relative 
contribution (in \%) of bending energy
 ({\it curves (1)}) and twist energy  ({\it
curves (2)}) to the total elastic energy stored in a
linker (within the chromatin superhelix), versus linker length (in bp),
  in various instances:
 ({\it above})
when  a force $\vec{F}=F\vec{A}$ is applied along the SH axis$\vec{A}$
but without any torque, ({\it middle}) 
when a torsional torque  $\vec{M}_t=M_t\vec{A}$
is applied but no force, ({\it below}) 
when a flexural torque ( $\vec{M}_b.\vec{A}=0$)
is applied but no force.
Two values of the angle $\Phi$ have been investigated:
$\Phi=90^o$ ({\it left}) or $\Phi=50^o$ ({\it right}).
The almost flat line, below 2 \%,  in the two upper figures,
 corresponds to the fraction of energy 
corresponding to stretch deformations; it is obviously negligible.

We compare these curves with  those obtained for an helical coiling
of DNA (section V.B) having the same values of $z$ and $r$
 ({\it crosses, curve (3)}
for the twist energy and {\it diamonds, curve (4)} for the bending energy).
When applying only a pulling force,
 the partitions 
 strongly differ;
by contrast, the partitions observed when only a torque
 (either $M_t$, either $\vec{M}_b$) is
applied are exactly identical (the curves superimpose).

\vskip 5mm\noindent
{\bf  Figure 13:}
Twist persistence length ${\cal C}$ ({\it dashed line, curve (1)}),
bend persistence length
${\cal A}$ ({\it solid line, curve (2)}) in nm of an helical spring
versus linker length (in bp)
with the values of $z$ and $r$ equal to those of the SH, respectively for
 $\Phi=90^o$ (left) and   $\Phi=50^o$ (right).
This shows that the decrease of the persistence lengths and their vanishing in special values
$n_{bp,c}$ is a geometric effect: it is due to an
accumulation of turns on a small distance along the axis, allowing a dramatic unfolding.

\vskip 5mm\noindent
{\bf  Figure 14:} Ribbon-like structure of the SH ($\theta=\pi$), here projected in the plane
($\vec{D}, \vec{N}\wedge\vec{D}$); indeed, when $\theta=\pi$, all the planes
($\vec{D}_j, \vec{N}_j\wedge\vec{D}_j$) coincide. The SH axis makes an 
angle $(\pi/2-\beta)$
quite small with this plane 
($\pi/2-\beta=\alpha\approx 4.47^o$) the nucleosomes are almost orthogonal to $\vec{A}$.
Obviously,  $D\sin\beta=l\sin(\Phi/2)$.

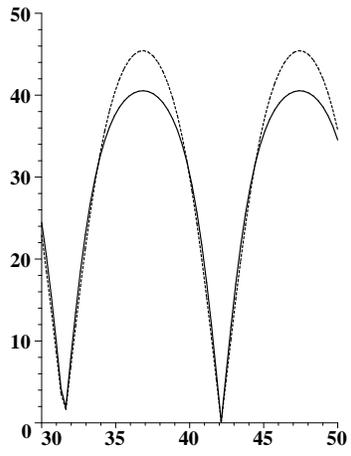
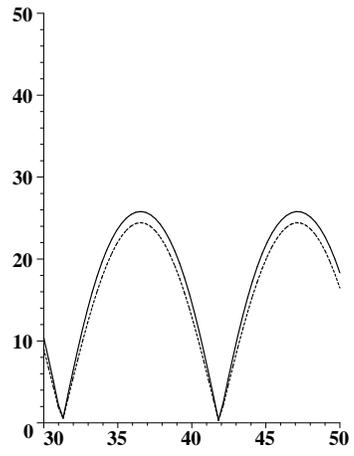
\end{document}